\documentclass[12 pt,  color]{article}
\usepackage{amsfonts, amsmath, amssymb, amsgen, amsthm, amscd,latexsym,mathrsfs}
\usepackage[all,cmtip]{xy}
\usepackage[latin5]{inputenc}
\usepackage{paralist}

\begin{document}

\begin{center}
\bigskip

{\Large On Geometry of Schmidt Legendre Transformation}

\bigskip

\bigskip

Oğul Esen\footnote{%
E-mail: oesen@gtu.edu.tr}

Department of Mathematics,

Gebze Technical University,

41400 Çay\i rova, Gebze, Kocaeli, Turkey,

\bigskip

\bigskip

Partha Guha\footnote{%
E-mail: partha@bose.res.in}

S.N. Bose National Centre for Basic Sciences

JD Block, Sector III, Salt Lake

Kolkata - 700098, India

\bigskip
\end{center}

\textbf{Abstract:} A geometrization of Schmidt-Legendre transformation of
the higher order Lagrangians is proposed by building a proper Tulczyjew's
triplet. The symplectic relation between Ostrogradsky-Legendre and
Schmidt-Legendre transformations is obtained. Several examples are presented.


\tableofcontents

\section{Introduction}

The dynamics of a system can either be formulated by a Lagrangian function
on the tangent bundle of a configuration space or by a Hamiltonian function
on the cotangent bundle \cite{AbMa78,Ar13}. For a physical system, ideally,
these two formalisms should be transformed to each other with the Legendre
transformations. The transformation is immediate by means of fiber
derivative of the Lagrangian function if Hessian of the Lagrangian (or
Hamiltonian) function is non-degenerate. If the Hessian is degenerate or/and
the system possesses various constraints, then defining the transformation
becomes complicated.

To overcome the obstructions due to the existences of singularities and
constraints, in the beginnings of the 50's, Dirac proposed an algorithm,
nowadays called Dirac-Bergmann algorithm \cite{Dirac,Dirac1950}. This
algorithm was geometrized at the end of 70's by Gotay, Nester and Hinds \cite%
{GoNe79,GoNe80,GoNe84,GoNeHi}. At the late 70's, Tulczyjew showed that the
dynamics can be represented as a Lagrangian submanifold of certain
symplectic manifold on higher order bundles \cite{Be11,Tu76,Tu77,TuUr99,We77}%
. In this setting, Hamiltonian and Lagrangian formulations are, in fact, two
different generators of the same Lagrangian submanifold. So that, Tulczyjew
redefined the Legendre transformation as a passage between these two
different generators.

Although in classical mechanics a Lagrangian density is a function of
positions and velocities, it is possible to find theories involving
Lagrangian densities depending on the higher order derivatives as well. In
such cases, to pass the Hamiltonian picture, it is a tradition to employ the
Ostrogradsky-Legendre transformation \cite{Ostra}. The Ostrogradsky approach
is based on the idea that consecutive time derivatives of initial
coordinates form new coordinates, hence a higher order Lagrangian can be
written in a form of a first order Lagrangian on a proper iterated tangent
bundle. In some recent studies, the higher order Lagrangians has received
attention \cite{BoKo05,Ma07,MaDa05,Mas16,Mas16b,Mo10,st06}, extending the
previous works of Pais and Uhlenbeck \cite{pu1950}. The linear harmonic
oscillator is a perennial favorite of quantum theorists besides being the
fundamental bedrock of many classical theoretical models in both physics and
engineering. In view of its fundamental importance it provides a good
reference point on which to build higher-order equations which have possible
physical relevance. There are extensive studies in the literature for
singular or/and constraint higher order Lagrangians as well, let us give an
incomplete list of such studies \cite%
{BaGoPoRo88,CrCa86,ChFaLiTo13,Fe96,KhRa96,GoRa94,GrPoRo91,NaHa96,Ne89,Po89}.
For the system whose configuration is a Lie group, we additionally refer
\cite{CoDe11,GaHoMeRa12,GaHoRa11}, and for field theories, see \cite{Vi09}.
We cite a recent study on the geometry of higher order theories in terms of
Tulczyjew's triplet \cite{CoPr16}.

At the middle of 90's, Schmidt proposed an alternative method for the
Legendre transformations of higher order Lagrangian systems working both for
non-degenerate and degenerate systems \cite{Sc94,Sc95}. Schmidt defined the
acceleration as a new coordinate instead of the velocity. Although,
Schmidt-Legendre transformation has not been credited as it deserves, it is
possible to find some related works in the literature \cite%
{DeSeYo09,HaHe02,Ka97}. Two works \cite{AnGoMaMa10,AnGoMa07} done on the
Schmidt-Legendre transformation have motivational importance for the present
work. In these papers, the comparison and the relation between the Legendre
transformations in the senses of Ostrogradsky and Schmidt have been studied.

The main concern of this manuscript is the higher order differential
equations generated by the higher order Lagrangian functions and their
Legendre transformations. The main objectives are to construct a geometric
framework, namely a Tulczyjew' triplet, for the Schmidt-Legendre
transformation, and construct the symplectic relations between
Ostrogradsky-Legendre and Schmidt-Legendre transformations in pure geometric
ways.

In order to achieve the goals of the paper, we shall start to the following
section by reviewing some basic ingredients of the Tulczyjew's construction
of the Legendre transformation, see also \cite{EsGu15a,EsGu15b}. In section $%
3$, we will review the geometry of higher order tangent bundles, the theory
of higher order Lagrangian formalism, and the Ostrogradsky-Legendre
transformation. In section $4$, the acceleration bundle will be defined.
Tangent and cotangent bundles of the acceleration bundle are presented.
Then, the Schmidt-Legendre transformation will be presented for the
Lagrangians in terms of the Tulczyjew triplet. I section $5$, the symplectic
relation between Ostrogradsky-Legendre and Schmidt-Legendre transformations
will be constructed. The last section, will be reserved for several examples
including Pais-Uhlenberg, Sar{\i}oğlu-Tekin and Clem\'ent Lagrangians.

\section{Generating Families}

\subsection{Special symplectic structures}

Let $P$ be a symplectic manifold carrying an exact symplectic two form $%
\Omega_{P}=d\vartheta_{P}$. A special symplectic structure is a quintuple $%
(P,\pi_{\mathcal{M}}^{P},\mathcal{M},\vartheta_{P},\chi)$ where $\pi_{%
\mathcal{M}}^{P}:P\rightarrow\mathcal{M}$ is a fibre bundle and $%
\chi:P\rightarrow T^{\ast}\mathcal{M}$ is a fiber preserving symplectic
diffeomorphism such that $\chi^{\ast}\theta_{T^{\ast}\mathcal{M}%
}=\vartheta_{P}$ for $\theta_{T^{\ast}\mathcal{M}}$ being the canonical
one-form on $T^{\ast}\mathcal{M}$. $\chi$ can be characterized uniquely by
the condition
\begin{equation*}
\left\langle \chi(p),X^{\mathcal{M}}(x)\right\rangle =\left\langle
\vartheta_{P}(p),X^{P}(p)\right\rangle
\end{equation*}
for each $p\in P$, $\pi_{M}^{P}(p)=x$ and for vector fields $X^{\mathcal{M}}$
and $X^{P}$ satisfying $\left( \pi_{\mathcal{M}}^{P}\right) _{\ast}X^{P}=X^{%
\mathcal{M}} $ \cite{LaSnTu75, SnTu73, Tu80}.

A real valued function $F$ on the base manifold $\mathcal{M}\mathbb{\ }$%
defines a Lagrangian submanifold
\begin{equation}
\mathcal{S}_{P}=\left\{ p\in P:d\left( F\circ\pi_{\mathcal{M}}^{P}\right)
(p)=\vartheta_{P}\left( p\right) \right\}  \label{LagSubSSS}
\end{equation}
of the underlying symplectic manifold $(P,\Omega_{P}=d\vartheta_{P})$. The
function $F$ together with a special symplectic structure $(P,\pi _{\mathcal{%
M}}^{P},\mathcal{M},\vartheta_{P},\chi)$ are called a generating family for
the Lagrangian submanifold $\mathcal{S}_{P}$. Since $\chi$ is a symplectic
diffeomorphism, it maps $\mathcal{S}_{P}$ to the image space $im\left(
dF\right) $ of the exterior derivative of $F$, which is a Lagrangian
submanifold of $T^{\ast}\mathcal{M}$.

\subsection{Morse Families}

Let $\left( P,\pi_{\mathcal{M}}^{P},\mathcal{M}\right) $ be a fibre bundle.
The vertical bundle $VP$ over $P$ is the space of vertical vectors $U\in TP$
satisfying $T\pi_{\mathcal{M}}^{P}\left( U\right) =0$. The conormal bundle
of $VP$ is defined by
\begin{equation*}
V^{0}P=\left\{ \alpha\in T^{\ast}P:\left\langle \alpha,U\right\rangle
=0,\forall U\in VP\right\} .
\end{equation*}
Let $E$ be a real-valued function on $P$, then the image $im\left( dE\right)
$ of its exterior derivative is a subspace of $T^{\ast}P $. We say that $E$
is a Morse family (or an energy function) if
\begin{equation}
T_{z}im\left( dE\right) +T_{z}V^{0}P=TT^{\ast}P,  \label{MorseReq}
\end{equation}
for all $z\in im\left( dE\right) \cap V^{0}P$, \cite{Be11, LiMa, Tu80,
TuUr96, TuUr99, We77}. In local coordinates $\left( x^{a},r^{i}\right) $ on
the total space $P$ induced from the coordinates $\left( x^{a}\right) $ on $%
\mathcal{M}$, the requirement in Eq.(\ref{MorseReq}) reduces to the
condition that the rank of the matrix%
\begin{equation}  \label{MF}
\left( \frac{\partial^{2}E}{\partial x^{a}\partial x^{b}}\text{ \ \ }\frac{%
\partial^{2}E}{\partial x^{a}\partial r^{i}}\right)
\end{equation}
be maximal. A Morse family $E$ on the smooth bundle $\left( P,\pi _{\mathcal{%
M}}^{P},\mathcal{M}\right) $ generates an immersed Lagrangian submanifold
\begin{equation}
\mathcal{S}_{T^{\ast}\mathcal{M}}=\left\{ \lambda_{M}\in T^{\ast}\mathcal{M}%
:T^{\ast}\pi_{\mathcal{M}}^{P}(\lambda_{\mathcal{M}})=dE\left( p\right)
\right\}  \label{LagSub}
\end{equation}
of $\left( T^{\ast}\mathcal{M},\Omega_{T^{\ast}\mathcal{M}}\right) $. Note
that, in the definition of $\mathcal{S}_{T^{\ast}\mathcal{M}}$, there is an
intrinsic requirement that $\pi_{\mathcal{M}}^{P}\left( p\right)
=\pi_{T^{\ast}\mathcal{M}}\left( \lambda_{\mathcal{M}}\right) $.

Consider a Morse family $E$ on a fibre bundle $P\rightarrow\mathcal{M}$, and
the Lagrangian submanifolds $\mathcal{S}_{T^{\ast}\mathcal{M}}$ defined in
Eq.(\ref{LagSub}). Assume that there exists a subbundle of $P^{\prime
}\subset P\rightarrow\mathcal{M}$ of where the induced function $%
E|_P^{\prime }$ satisfies the conditions of being a Morse family presented
in (\ref{MF}) and generates the same Lagrangian submanifold $\mathcal{S}%
_{T^{\ast}\mathcal{M}}$ of $T^{\ast}\mathcal{M}$. This procedure is called
the reduction of the Morse family. Note that, in this case, the final
structure $E|_{P^{\prime }}$ on $P^{\prime }\rightarrow\mathcal{M}$ is
called the reduced Morse family generating $\mathcal{S}_{T^{\ast}\mathcal{M}}
$. We refer \cite{Be11,TuUr07} for more elaborated formulations and more
precise discussions on this subject.

\subsection{The Legendre transformation}

Let $\left( P,\Omega_{P}=d\vartheta_{P}\right) $ be an exact symplectic
manifold, and $(P,\pi_{\mathcal{M}}^{P},\mathcal{M},\vartheta_{P},\chi)$ be
a special symplectic structure. A function $F$ on $\mathcal{M}$ defines a
Lagrangian submanifold $\mathcal{S}_{P}\subset P$ as described in Eq.(\ref%
{LagSubSSS}). If $\mathcal{S}_{P}=im\left( \Upsilon\right) $ is the image of
a section $\Upsilon$ of $\left( P,\pi_{\mathcal{M}}^{P},\mathcal{M}\right) $
then we have $\chi\circ\Upsilon=dF$. Assume that $(P,\pi_{\mathcal{M}%
^{\prime}}^{P},\mathcal{M}^{\prime},\vartheta_{P}^{\prime },\chi^{\prime})$
is another special symplectic structure associated to the underlying
symplectic space $\left( P,\Omega_{P}\right) $. Then, from the diagram%
\begin{equation}
\xymatrix{T^{\ast }\mathcal{M} \ar[dr]^{\pi_{\mathcal{M}}} &&P
\ar[dl]^{\pi^{P}_{\mathcal{M}}} \ar[rr]^{\chi^{\prime }}
\ar[dr]^{\pi^{P}_{\mathcal{M}^{\prime }}} \ar[ll]_{\chi}&&T^{\ast
}\mathcal{M}^{\prime } \ar@<1ex>[dl]^{\pi_{\mathcal{M}^{\prime }}}
\\&\mathcal{M} \ar@<1ex>[ul]^{dF} \ar@<1ex>[ur]^{\Upsilon}
&&\mathcal{M}^{\prime} \ar[ur]^{dF^{\prime }}
\ar@<1ex>[ul]^{\Upsilon^{\prime}} }  \label{TTG}
\end{equation}
it follows that the difference $\vartheta_{P}-\vartheta_{P}^{\prime}$ of
one-forms must be closed in order to satisfy $\Omega_{P}=d\vartheta
_{P}=d\vartheta_{P}^{\prime}$. When the difference is exact, there exists a
function $\Delta$ on $P$ satisfying $d\Delta=\vartheta_{P}-\vartheta
_{P}^{\prime}$. If $\mathcal{S}_{P}$ is the image of a section $\Upsilon
^{\prime}$ of the fibration $\left( P,\pi_{\mathcal{M}^{\prime}}^{P},%
\mathcal{M}^{\prime}\right) $, then the function
\begin{equation}
F^{\prime}=\left( F\circ\pi_{\mathcal{M}}^{P}+\Delta\right) \circ
\Upsilon^{\prime}
\end{equation}
generates the Lagrangian submanifold $\mathcal{S}_{P}$ \cite{Tu77, TuUr96,
TuUr99}. This is the Legendre transformation. If, finding a global section $%
\Upsilon^{\prime}$ of $\pi_{\mathcal{M}^{\prime}}^{P}$ satisfying $%
im(\Upsilon^{\prime})=\mathcal{S}_{P}$ is not possible, the Legendre
transformation is not immediate. In this case, define the Morse family
\begin{equation}
E=F\circ\pi_{\mathcal{M}}^{P}+\Delta  \label{energy}
\end{equation}
on a smooth subbundle of $\left( P,\pi_{\mathcal{M}^{\prime}}^{P},\mathcal{M}%
^{\prime}\right) $, where $E$ satisfies the requirement (\ref{MorseReq}) of
being a Morse family. Then, $E$ generates a Lagrangian submanifold $\mathcal{%
S}_{T^{\ast}\mathcal{M}^{\prime}}$ on $T^{\ast }\mathcal{M}^{\prime}$ as
described in Eq.(\ref{LagSub}). The inverse of $\chi^{\prime}$ maps $%
\mathcal{S}_{T^{\ast}\mathcal{M}^{\prime}}$ to $\mathcal{S}_{P}$
bijectively, that is $\mathcal{S}_{P}=\left( \chi^{\prime }\right)
^{-1}\left( \mathcal{S}_{T^{\ast}\mathcal{M}^{\prime}}\right) $.

A way to generate the symplectic transformation between $T^{\ast }\mathcal{M}
$ and $T^{\ast }\mathcal{M}^{\prime }$ that is generate a symplectic
diffeomorphism $\chi^{\prime -1}$ e.g. from $T^{\ast }\mathcal{M}$ to $%
T^{\ast }\mathcal{M}^{\prime }$ is to define a generating function $F$ on $%
\mathcal{M}\times \mathcal{M}^{\prime }$. This transforms the Lagrangian
submanifold of $T^*\mathcal{M}$ defined by the Morse family $F$ on $\mathcal{%
M}\times \mathcal{M}^{\prime }$ over $\mathcal{M}$ to the the Lagrangian
submanifold of $T^*\mathcal{M}^{\prime }$ defined by the Morse family $F$ on
$\mathcal{M}\times \mathcal{M}^{\prime }$ over $\mathcal{M}^{\prime }$.

\section{Ostrogradsky-Legendre transformation}

\subsection{Higher order tangent bundles}

Let $Q$ be an n-dimensional differentiable manifold with coordinates $%
x\rightarrow\left( \mathbf{q}\right) =(q^{1},q^{2},...,q^{n})$. Two
differentiable curves $\gamma$ and $\tilde{\gamma}$ are called equivalent $%
\gamma\sim_{k}\hat{\gamma}$ if they agree at $x$ and agree up to their $k-$%
th derivatives, that is if
\begin{equation}
\left. \frac{d^{r}}{dt^{r}}\right\vert _{t=0}\gamma=\left. \frac{d^{r}}{%
dt^{r}}\right\vert _{t=0}\hat{\gamma}\text{ \ \ for }r=0,1,2,...,k.
\label{ERk}
\end{equation}
An $k$-th order tangent vector $v^{k}\left( x\right) $ at $x$ is an
equivalence class of curves at $x.$ This class will be denoted by $%
t^{k}\gamma\left( 0\right) $. The set of all equivalence classes of curves,
that is the set of all $k$-th order tangent vector at $x$ is the $k$-th
order tangent space $T_{x}^{k}Q$ at $x\in Q.$ Let $\gamma$ be a
representative of $v_{x}^{k}\in T_{x}^{k}Q,$ then
\begin{equation*}
v_{x}^{k}\rightarrow(\mathbf{q,\dot{q},\ddot{q},...,q}^{(k)}).
\end{equation*}

The sum $T^{k}Q=\cup T_{x}^{k}Q$ of all $k$-th order tangent spaces $%
T_{x}^{k}Q$ as $x$ varies on $Q$ is the total space of the $k$-th order
tangent bundle of $Q$ with fibers being $T_{x}^{k}Q$. The $k-th$ order
tangent bundle $T^{k}Q$ of $Q$ is $\left( k+1\right) n$ dimensional
manifold. There are hierarchic fibrations defined on each fiber
\begin{equation}
_{s}^{r}\tau_{Q}:T_{x}^{r}Q\rightarrow T_{x}^{s}Q:t^{r}\gamma(0)\rightarrow
t^{s}\gamma(0)  \label{tbp}
\end{equation}
where $r>s$, and $r=1,...,k$ and $s=0,1,...,k-1$. In local charts, this
looks like projecting an $r+1$-tuple $(\mathbf{q;\dot{q};\ddot{q};...;q}%
^{(r)})$ to its first $s+1$ components $(\mathbf{q;\dot{q};\ddot{q};...;q}%
^{(s)})$.

\subsection{Higher order Euler-Lagrange equations}

Let $Q$ be an $n$-dimensional manifold with local coordinates $\left(
\mathbf{q}\right) =(q^{1},q^{2},...,q^{n})$. The $k-th$ order tangent bundle
$T^{k}Q$ of $Q$ is $\left( k+1\right) n$ dimensional manifold with local
coordinates $(\mathbf{q;\dot{q};\ddot{q};...;q}^{(k)})$ induced those from $%
Q $. A Lagrangian density $L=L(\mathbf{q;\dot{q};\ddot{q};...;q}^{(k)})$ is
a real-valued function on $T^{k}Q$ from which we define the action integral
\begin{equation*}
\mathcal{S}=\int_{a}^{b}{L}(\mathbf{q;\dot{q};\ddot{q};...;q}^{(k)})dt
\end{equation*}
by fixing two points $\mathbf{q}\left( a\right) $ and $\mathbf{q}\left(
b\right) $ in $Q$. To find the extremum values, we take the variation of the
action integral which results with $k$-th order Euler-Lagrange equations%
\begin{equation}
\sum_{\alpha=0}^{k}(-1)^{\alpha}\frac{d^{\alpha}}{dt^{\alpha}}\left( \frac{%
\partial L}{\partial\mathbf{q}^{(\alpha)}}\right) =\mathbf{0}.  \label{ELa}
\end{equation}
If the partial derivative $\partial L/\partial\mathbf{q}^{(k)}$ depends on $%
\mathbf{q}^{(k)}$ then, $k$-th order Euler-Lagrange equations (\ref{ELa})
are a set of differential equation of order $2k$.

\subsection{Ostrogradsky-Legendre transformation}

The traditional framework for obtaining the Hamiltonian formulation of the
higher order Lagrangian formalisms is due to Ostrogradsky \cite{Ostra}. The
Ostrogradsky approach is based on the idea that consecutive time derivatives
of initial coordinates form new coordinates, hence a higher order Lagrangian
can be written in a form of a first order Lagrangian on iterated tangent
bundles.

There are several different ways to express a $k$-th order Lagrangian
formalism in a first order form. Most general one is to define $(k+1)n$%
-dimensional configuration space $N=T^{k}Q$ with coordinates
\begin{equation*}
\mathbf{q}_{\left( 1\right) }=\mathbf{q},\qquad \mathbf{q}_{\left( 2\right)
}=\mathbf{\dot{q}},\qquad ...,\qquad \mathbf{q}_{\left( k\right) }=\mathbf{q}%
^{(k-1)}
\end{equation*}
by imposing the set of constraints $\mathbf{\dot{q}}_{\left( i\right) }-%
\mathbf{q}_{\left( i+1\right) }=0$ for $i=1,...,k-1$. In this case, starting
with a $k$-th order Lagrangian $L=L\left( \mathbf{q;\dot{q};...;q}%
^{(k)}\right) $ on $T^{k}Q$, we define the following constraint Lagrangian
density%
\begin{equation}
L_{N}={L}\left( \mathbf{q}_{\left( 1\right) },\mathbf{q}_{\left( 2\right)
},...,\mathbf{q}_{\left( k\right) }, \mathbf{\dot{q}}_{\left( k\right) }
\right) +\mathbf{\lambda}_1\cdot\left( \mathbf{\dot{q}}_{\left( 1\right) }-%
\mathbf{q}_{\left( 2\right) }\right)+...+ \mathbf{\lambda}_{k-1}\cdot\left(
\mathbf{\dot{q}}_{\left( k-1\right) }-\mathbf{q}_{\left( k\right) }\right)
\label{L_N}
\end{equation}
on the tangent bundle $T(T^{k}Q\times\mathbb{R}^{(k-1)n})$ with base
coordinates $\left( \mathbf{q}_{\left( 1\right) },...,\mathbf{q}_{\left(
k\right) },\lambda_1,...,\lambda_{k-1},\right) $.
In this formulation, the canonical Hamiltonian function is given by
\begin{equation}
H=\sum_{i=1}^{n}\mathbf{\dot{q}}_{\left( i\right) }\cdot\mathbf{\pi}^{i}-{L}%
\left( \mathbf{q}_{\left( 1\right) },\mathbf{q}_{\left( 2\right) },...,%
\mathbf{q}_{\left( k\right) }, \mathbf{\dot{q}}_{\left( k\right) } \right),
\label{canH2}
\end{equation}
where $(\mathbf{\pi}^{1},...,\mathbf{\pi}^{k})$ is set of the conjugate
momenta and the local representatives of the cotangent bundle $T^*N$. Note
that, $\mathbf{\dot{q}}_{\left( k\right) }$ is a function of $\left( \mathbf{%
q}_{\left( 1\right) },...,\mathbf{q}_{\left( k\right) },\mathbf{\pi }%
^{k}\right) $ if the matrix $[\partial^{2}L/\partial\mathbf{q}_{(k)}^2]$ is
non degenerate.

This motivates a shorter way of defining momenta without refereing to the
Lagrangian multipliers as follows. The Ostrogradsky -Legendry transformation
can be realised by the introduction of the momenta
\begin{equation}
T_qN\rightarrow T^*_qN:(\mathbf{\dot{q}}_{\left( 1\right) },\mathbf{\dot{q}}%
_{\left( 2\right) },...,\mathbf{\dot{q}}_{\left( k\right) } )\rightarrow(%
\mathbf{\pi}^{1},...,\mathbf{\pi}^{k}),  \label{OsMo}
\end{equation}
where
\begin{equation}
\pi^k=\sum_{j=k}^n\left(-\frac{d}{dt}\right)^{j-k}\left(\frac{\partial L}{%
\partial\mathbf{q}^{(j)}}\right)
\end{equation}
see also for the step by step transformation \cite{BaGoPoRo86,
BaGoPoRo88,GrPoRo91}.

\section{Schmidt-Legendre transformation}

\subsection{Acceleration bundle}

Let $x$ be a point in a manifold $Q$, and consider the set $%
C_{x}^{\infty}\left( Q\right)$ of smooth curves passing through $x$. We
define a subset $A_x\left( Q\right)$ of $C_{x}^{\infty}\left( Q\right)$ by
only considering the curves whose first derivatives are vanishing at $x$.
That is, if $\gamma$ is curve in $A_x\left( Q\right)$, then $\gamma(0)=x$
and $\dot{\gamma}(0)=\mathbf{0}\in T_x Q$. It is worthless to say that since
the vanishing of the first derivative is asked only at a single point, the
curve $\gamma$ needs not to be a constant.

We now define an equivalence relationship on $A_x\left( Q\right)$. Take two
curves $\gamma,\beta\in A_x\left( Q\right)$, that is $\gamma\left( 0\right)
=\beta\left( 0\right)=x$ and $\dot{\gamma}\left( 0\right) =\dot{\beta}(0)=%
\mathbf{0}$. We call $\gamma$ and $\beta$ are equivalent if they are agree
up to their second derivatives. In other words, two equivalent curves $\gamma
$ and $\beta$ satisfy
\begin{equation*}
\gamma\left( 0\right) =\beta\left( 0\right)=x, \qquad \dot{\gamma}\left(
0\right) =\dot{\beta}(0)=\mathbf{0}, \qquad \ddot{\gamma}\left( 0\right) =%
\overset{\cdot\cdot}{\beta}\left( 0\right) .
\end{equation*}
We denote an equivalence class by $t^{2}\gamma\left( 0\right)$ and denote it
by $a\left( x\right)$. The set of all equivalence classes of curves at $x$
is the acceleration space $A_{x}Q$ at $x\in Q$. We denote an equivalence
class $t^{2}\gamma\left( 0\right)$ by $a_{x}\in A_{x}Q$. The union $AQ=\cup
A_{x}Q$ of all acceleration spaces is a $2n-$dimensional manifold with local
charts $a_{x}\rightarrow(\mathbf{q,a=\ddot{q}})$ induced those from the
coordinates on $Q$. This suggests the fiber bundle structure of the
acceleration manifold $ AQ$ over $Q$ with projection
\begin{equation*}
\mathfrak{t}_{Q}: AQ\rightarrow Q:a_{x}\rightarrow x.
\end{equation*}

An alternative definition of the acceleration bundle may be stated as
follows. First recall the imbedding of the second order tangent bundle $%
T^{2}Q$ into the iterated bundle $TTQ:=T(TQ)$. An element of $V\in TTQ$ is
in the image space of the imbedding if it satisfies the equality $%
T\tau_{Q}(V)=\tau_{TQ}(V)$, where $T\tau_{Q}$ is the tangent lift of the
projection $\tau_{Q}:TQ\rightarrow Q$, and $\tau_{TQ}$ is the tangent bundle
projection $TTQ\rightarrow TQ$. In literature, such an element $V$ is called
second order. To define an imbedding of the acceleration bundle $AQ$ into $%
TTQ$, we additionally require that a second order vector field must satisfy $%
T\tau_{Q}(V)=\tau_{TQ}(V)=\mathbf{0}$. By this way, we define $AQ$ as a
subbundle of the bundle $TTQ\rightarrow Q$. In this respect, we may also
understand $AQ$ as a subbundle of $T^{2}Q\rightarrow Q$ as well.
Diagrammatically, we summarize the sequence of subbundles as follows.
\begin{equation}
\begin{array}{ccccc}
\begin{array}{c}
AQ \\
\left( \mathbf{q,a}\right)%
\end{array}
& \rightarrow &
\begin{array}{c}
T^{2}Q \\
\left( \mathbf{q,0,a}\right)%
\end{array}
& \rightarrow &
\begin{array}{c}
TTQ \\
\left( \mathbf{q,0;0,a}\right)%
\end{array}%
\end{array}
\label{dia1}
\end{equation}
This gives that, the acceleration bundle $AQ$ can be identified with the
intersection of the vertical subbundle $VTQ$ (consisting of vectors on $TQ$
and projecting to the zero vector on $Q$ via the mapping $T\tau_{Q}$) of $TTQ
$ and the second order tangent bundle $T^{2}Q$. Alternatively, we may write
a sequence
\begin{equation*}
AQ\rightarrow T^{2}Q\rightarrow TQ,
\end{equation*}
where the first mapping is the inclusion and the second is the projection $%
_{1}^{2}\tau_{Q}$ in (\ref{tbp}). When a connection is defined on $Q$, $%
T^{2}Q$ becomes a vector bundle over $Q$, and it can be written as the
Withney some of two copies of its tangent bundle \cite{DoRa80}. For a more
general discussion on this, we refer \cite{Su13}. In such a case, one may
identify the acceleration bundle $AQ$ with the tangent bundle $TQ$.

Let $\phi$ be a differential mapping $Q\rightarrow M$ between two manifolds $%
Q$ and $M$. We define the acceleration lift $A\phi: AQ\rightarrow  AM$ of $%
\phi$ as follows. Take $a_{x}\in A_{x}Q$, it can be represented by a curve $%
\gamma$ lying in the equivalence class $a_{x}=t^{2}\gamma\left( 0\right)$.
The image space of the curve under $\phi\circ \gamma$ is a curve in $M$.
Note also that, the velocity of $\phi\circ \gamma$ at $t=0$ vanishes, hence $%
\phi\circ \gamma$ is lying in one of the equivalence classes in the
acceleration bundle $A_{\phi{(x)}}M$ consisting of the velocity free curves
passing through $\phi{(x)}\in M$. Accordingly, we define the acceleration
lift by
\begin{equation}  \label{acclift}
A\phi\left( t^{2}\gamma\left( 0\right)\right) =t^{2}\left(
\phi\circ\gamma\right) ( 0)
\end{equation}
where $t^{2}\left(\phi\circ\gamma\right) ( 0)$ is the equivalence class
containing $\phi\circ \gamma$.

Now, we define a local diffeomorphism linking the tangent and acceleration
bundles and second order tangent bundle. Let a curve $\gamma$ represents an
element in $T^2 _x Q$, then one may write $\gamma(t)=x+vt+at^2+O(t)$. We
define a curve $\tilde{\gamma}(t)=x+at^2+O(t)$ which is velocity free.
Consider the following mapping
\begin{equation}
T^2 _x Q\rightarrow A _x Q \times T_x Q : t^2\gamma(0) \rightarrow (t^2%
\tilde{\gamma}(0),t\gamma(0))
\end{equation}
In a local coordinate frame, this mapping looks like
\begin{equation*}
T^2 Q\rightarrow A Q \times T Q : (q,\dot{q},a) \rightarrow (q,\dot{q},a).
\end{equation*}

It is possible to define the mixed iterated tangent and acceleration bundles
$ATQ$ and $T AQ$ with local coordinate charts
\begin{align*}
(\mathbf{q,\dot{q};a}_{q},\mathbf{a}_{\dot{q}}) & : ATQ\rightarrow\mathbb{R}%
^{4n} \\
(\mathbf{q,a;\dot{q}},\mathbf{\dot{a}}) & :T AQ\rightarrow\mathbb{R}^{4n}.
\end{align*}
It is evident that, there exist various projections of the iterated bundles
given by
\begin{align}
\tau_{ AQ} & :T AQ\rightarrow AQ:(\mathbf{q,a;\dot{q}},\mathbf{\dot{a}}%
)\rightarrow\left( \mathbf{q,a}\right) ,  \notag \\
T\mathfrak{t}_{Q} & :T AQ\rightarrow TQ:(\mathbf{q,a;\dot{q}},\mathbf{\dot{a}%
})\rightarrow\left( \mathbf{q,\dot{q}}\right) ,  \notag \\
\mathfrak{t}_{TQ} & : ATQ\rightarrow TQ:(\mathbf{q,\dot{q};a}_{q},\mathbf{a}%
_{\dot{q}})\rightarrow\left( \mathbf{q,\dot{q}}\right) ,  \label{accT} \\
A\tau_{Q} & : ATQ\rightarrow AQ:(\mathbf{q,\dot {q};a}_{q},\mathbf{a}_{\dot{q%
}})\rightarrow\left( \mathbf{q,a}_{q}\right).  \notag
\end{align}
Consider also the following velocity-acceleration rhombics establishing the
double fiber bundle structures of $ATQ$ and $T AQ$.
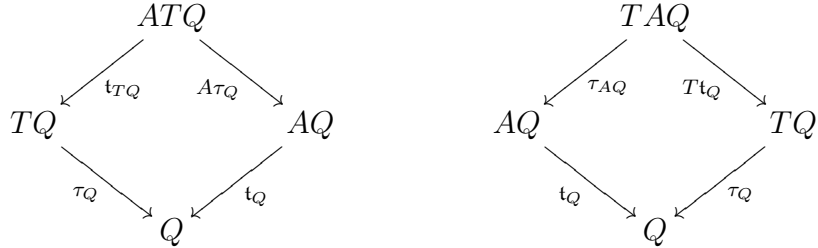
\begin{equation*}
\xymatrix{& ATQ\ar[dl]^{\mathfrak{t}_{TQ}}\ar[dr]_{ A\tau_{Q}} &&&&T
AQ\ar[dl]^{\tau_{ AQ}}\ar[dr]_{T\mathfrak{t}_{Q}} \\ TQ \ar[dr]_{\tau_Q} &&
AQ\ar[dl]^{\mathfrak{t}_{Q}} && AQ\ar[dr]_{\mathfrak{t}_{Q}}&&TQ
\ar[dl]^{\tau_Q} \\ &Q&&&&Q }
\end{equation*}
These commutative diagrams enable us to define the canonical maps%
\begin{align}
\kappa_{Q} & : ATQ\rightarrow T AQ:(\mathbf{q,\dot{q};a}_{q},\mathbf{a}_{%
\dot{q}})\rightarrow(\mathbf{q,\mathbf{a}_{q};\dot{q}},\mathbf{a}_{\dot{q}}),
\notag \\
\hat{\kappa}_{Q} & :T AQ\rightarrow ATQ:(\mathbf{q,\mathbf{a};\dot{q}},%
\mathbf{\mathbf{\dot{a}}})\rightarrow (\mathbf{q,\dot{q};\mathbf{a,\dot{a}}}%
).  \label{kappahat}
\end{align}

We note that, the third order tangent bundle $T^{3}Q$ and the acceleration
bundle $ATQ$ are isomorphic and composing this with the canonical map $%
\kappa_{Q}$ in (\ref{kappahat}), we arrive the bundle isomorphism $T
AQ\rightarrow T^{3}Q$ via the map
\begin{equation}
S:TAQ\rightarrow T^{3}Q:\left( \mathbf{q,a;\dot{q},\dot{a}}\right)
\rightarrow(\mathbf{q},\mathbf{\dot{q}};\mathbf{a},\mathbf{\dot{a}}).
\label{iso}
\end{equation}

\subsection{Higher order acceleration bundles}

Let us generalize the previous discussions on the acceleration bundle to the
bundles higher than $3$. To this end, first define curves on a base manifold
$Q$ with coordinates $(q)$ and consider the second order acceleration space $%
A^2_q Q$ at the point $q\in Q$ by defining an equivalence relationship on
the space of curves with vanishing velocities $\dot{q}(t)=0$ and the third
derivative $\dddot{q}(t)=0$ by requiring that their second and fourth
derivatives are equal. The collection of these spaces results with the
second order acceleration bundle $A^2Q=\bigcup_{q\in Q}A^2_q Q$. The
projection
\begin{equation*}
A^2Q\rightarrow Q:(\mathbf{q},\mathbf{a}_q,\mathbf{b}_q)\rightarrow(\mathbf{q%
})
\end{equation*}
defines a fiber bundle structure over $Q$. One may define the following
iterative picture as well
\begin{equation*}
A^2Q\rightarrow AQ \rightarrow Q:(\mathbf{q},\mathbf{a}_q,\mathbf{b}%
_q)\rightarrow(\mathbf{q},\mathbf{a}_q)\rightarrow(\mathbf{q}).
\end{equation*}
To generalize this to higher order bundles is straight forward. So that, one
can define $A^kQ$ and the following projections
\begin{equation*}
A^kQ\rightarrow A^{k-1}Q \rightarrow... \rightarrow AQ\rightarrow Q.
\end{equation*}

Let us now make the local isomorphism between the tangent bundle of a higher
order acceleration bundle with the odd order iterated tangent bundles as
follows. For the fifth order tangent bundle $T^5Q$, the isomorphism is given
by
\begin{equation*}
TA^2Q\rightarrow T^5Q: (\mathbf{q},\mathbf{a}_q,\mathbf{b}_q;\mathbf{\dot{q}}%
,\mathbf{\dot{a}}_q,\mathbf{\dot{b}}_q)\rightarrow (\mathbf{q};\mathbf{\dot{q%
}};\mathbf{a}_q;\mathbf{\dot{a}}_q;\mathbf{b}_q;\mathbf{\dot{b}}_q) .
\end{equation*}
In a similar we may generalized this by identifying
\begin{equation*}
TA^rQ\rightarrow T^{(2r+1)}Q: (\mathbf{q},\mathbf{a}_q,...,\mathbf{b}_q;%
\mathbf{\dot{q}},\mathbf{\dot{a}}_q,...,\mathbf{\dot{b}}_q)\rightarrow (%
\mathbf{q};\mathbf{\dot{q}};\mathbf{a}_q;\mathbf{\dot{a}}_q;...;\mathbf{b}_q;%
\mathbf{\dot{b}}_q) .
\end{equation*}

\subsection{Gauge symmetry of second order Lagrangian formalisms}

The essence of the Schmidt's Method presented in \cite%
{AnGoMaMa10,AnGoMa07,Sc94,Sc95} is to replace the higher-order
Euler-Lagrange equation of motion by the more familiar first-order
Euler-Lagrange equation by the introduction of a new Lagrangian. As it is
very well known, the symmetry of the Euler-Lagrange equations (\ref{ELa}) in
any order is the addition of a total derivative to the Lagrangian $L$.
Consider a second order non-degenerate Lagrangian density%
\begin{equation}
{L=L}\left( \mathbf{q;\mathbf{\dot{q};}\ddot{q}}\right)  \label{Lag}
\end{equation}
on $T^{2}Q$. We add the total derivative of a function $F=F\left( \mathbf{q,%
\dot{q},\ddot{q}}\right) $ in order to arrive a new Lagrangian function
\begin{align}
\hat{L}\left( \mathbf{q;\mathbf{\dot{q};}\ddot{q};\dddot{q}}\right) & ={L}%
\left( \mathbf{q;\mathbf{\dot{q};}\ddot{q}}\right) +\frac{d}{dt}F\left(
\mathbf{q;\dot{q};\ddot{q}}\right)  \notag \\
& ={L}\left( \mathbf{q;\mathbf{\dot{q};}\ddot{q}}\right) +\frac{\partial F}{%
\partial\mathbf{q}}\mathbf{\dot{q}}+\frac{\partial F}{\partial \mathbf{\dot{q%
}}}\mathbf{\ddot{q}+}\frac{\partial F}{\partial\mathbf{\ddot{q}}}\mathbf{%
\dddot{q}}  \label{Lhat}
\end{align}
of order $3$. Note that, $\hat{L}$ is defined on $T^{3}Q$ whereas ${L}$ and $%
F$ are defined on $T^{2}Q$. A straightforward computation proves that, the
third order Euler-Lagrange equations generated by $\hat{L}$ are the same
with the second order Euler-Lagrangian equations generated by ${L}$.

By recalling the surjection $S $ presented in Eq.(\ref{iso}), we pull the
Lagrangian $\hat{L}$ in Eq.(\ref{Lhat}) back to the tangent bundle $TAQ$ of
the acceleration bundle $AQ$ which results with a first order Lagrangian
\begin{equation}
L_{2}\left( \mathbf{Q},\mathbf{\mathbf{A};\mathbf{\dot{Q}},\mathbf{\dot{A}}}%
\right) ={L}\left( \mathbf{Q},\mathbf{\dot{Q}},\mathbf{\mathbf{A}}\right) +%
\frac{\partial F}{\partial\mathbf{Q}}\mathbf{\mathbf{\dot{Q}}+}\frac{%
\partial F}{\partial\mathbf{\mathbf{\dot{Q}}}}\mathbf{\mathbf{A}}+\frac{%
\partial F}{\partial\mathbf{\mathbf{A}}}\mathbf{\mathbf{\dot{A}}}  \label{L2}
\end{equation}
on the first order tangent bundle $TAQ$ considering the base manifold as $AQ$%
. The first order Euler-Lagrange equation generated by the Lagrangian
density (\ref{L2}) takes the particular form
\begin{equation}
\frac{\partial L_{2}}{\partial\mathbf{Q}}-\frac{d}{dt}\frac{\partial L_{2}}{%
\partial\mathbf{\dot{Q}}}=0\text{ \ \ and \ \ }\frac{\partial L_{2}}{\partial%
\mathbf{\mathbf{A}}}-\frac{d}{dt}\frac{\partial L_{2}}{\partial\mathbf{%
\mathbf{\dot{A}}}}=0.  \label{ELtwo}
\end{equation}
We find upon expanding the total time derivative in the second equation in (%
\ref{ELtwo}) and using the fact that $L$ is independent of $\mathbf{\mathbf{%
\dot{A}}}$
\begin{equation}
\left( \frac{\partial L}{\partial\mathbf{\mathbf{A}}}+\frac{\partial F}{%
\partial\mathbf{\dot{Q}}}\right) +\frac{\partial^{2}F}{\partial \mathbf{\dot{%
Q}}\partial\mathbf{\mathbf{A}}}\cdot(\mathbf{\mathbf{A}}-\mathbf{\mathbf{%
\ddot{Q}}})=\mathbf{0}.  \label{cons}
\end{equation}
Recall that, we have started with a second order non-degenerate Lagrangian
density $L$ presented in (\ref{Lag}). The non-degeneracy of $L$ is
equivalent to say that the rank of the Hessian matrix $[\partial^{2}L/%
\partial\mathbf{\mathbf{A}}^{2}]$ is full. We additionally assume that the
auxiliary function $F$ satisfies the equality
\begin{equation}
\frac{\partial L}{\partial\mathbf{\mathbf{A}}}+\frac{\partial F}{\partial%
\mathbf{\mathbf{\dot{Q}}}}=\mathbf{0}.  \label{chi}
\end{equation}
In this case, taking the partial derivative of this equality (\ref{chi})
with respect to $\mathbf{A}$, we observe that the matrix $%
[\partial^{2}F/\partial\mathbf{\mathbf{A}}\partial \mathbf{\dot{Q}}]$ is
non-degenerate. In this case, the second set of the Euler-Lagrangian
equations presented in (\ref{ELtwo}) reduce to the constraint $\mathbf{%
\mathbf{A}}-\mathbf{\mathbf{\ddot{Q}}=0}$. Accordingly, the first set of the
Euler-Lagrangian equations in (\ref{ELtwo}) give the second order
Euler-Lagrange equations generated by the Lagrangian $L$ in (\ref{Lag}).

Let us take the partial derivative of the assumption on the auxiliary
function $F$ in (\ref{chi}) with respect to $\mathbf{\dot{Q}}$. This brings
an integrability condition on $F$ as follows. The symmetry of the matrix $%
[\partial^{2}F/\partial\mathbf{\mathbf{\dot{Q}}}^{2}]$ enforces the symmetry
of the matrix $[\partial^{2}L/\partial\mathbf{\mathbf{A}}\partial \mathbf{%
\mathbf{\dot{Q}}]}$. In other words, to define a proper auxiliary function,
the matrix $[\partial ^{2}L/\partial\mathbf{\mathbf{A}}\partial\mathbf{%
\mathbf{\dot{Q}}]}$ has to be symmetric. If not, then there may be no such
auxiliary function $F$ establishing the reduction. In the forthcoming
subsection, we shall present some examples pointing out this fact.

\subsection{Schmidt-Legendre transformation for even orders}

This construction enables us to interpret the second order Euler-Lagrange
equations as Lagrangian submanifolds. To this end we propose the following
Tulczyjew's triplet for the case of acceleration bundle
\begin{equation}
\xymatrix{T^{\ast }T AQ \ar[dr]^{\pi_{T AQ}}&&TT^{\ast } AQ\ar[dl]^{T\pi_{
AQ}} \ar[rr]^{\Omega_{T^{\ast} AQ}^{\flat}} \ar[dr]_{\tau_{T^{\ast } AQ}}
\ar[ll]_{\alpha_{ AQ}}&&T^{\ast }T^{\ast }Q\ar@<1ex>[dl]^{\pi _{T^{\ast}
AQ}} \\&T AQ\ar@<1ex>[ul]^{dL_2} &&T^{\ast} AQ\ar[ur]^{-dH}}  \label{TAcc}
\end{equation}
where $L_{2}$ is the Lagrangian density in (\ref{L2}). Introducing the
coordinates
\begin{equation*}
( \mathbf{Q,A,P_Q,P_A;\dot{Q},\dot{A},\dot{P}_Q,\dot{P}_A})\in TT^{\ast}AQ
\end{equation*}
one finds the symplectomorphisms
\begin{eqnarray}
\alpha_{AQ}( \mathbf{Q,A,P_Q,P_A;\dot{Q},\dot{A},\dot{P}_Q,\dot{P}_A}) &=&(
\mathbf{Q,A,\dot{Q},\dot{A};\dot{P}_Q,\dot{P}_A,P_Q,P_A}) \\
\Omega_{T^{\ast}AQ}^{\flat }( \mathbf{Q,A,P_Q,P_A;\dot{Q},\dot{A},\dot{P}_Q,%
\dot{P}_A}) &=& ( \mathbf{Q,A,P_Q,P_A;-\dot{P}_Q,-\dot{P}_A,\dot{Q},\dot{A}})
\end{eqnarray}
and the potential one-forms
\begin{eqnarray}
\vartheta_{1}&=&\mathbf{\dot{P}}_Q\cdot d\mathbf{Q}+ \mathbf{\dot{P}}_A\cdot
d\mathbf{A}- \mathbf{\dot{Q}}\cdot d\mathbf{P_Q}-\mathbf{\dot{A}}\cdot d%
\mathbf{P_A}, \\
\vartheta_{2}&=&\mathbf{\dot{P}}_Q\cdot d\mathbf{Q}+ \mathbf{\dot{P}}_A\cdot
d\mathbf{A} +\mathbf{P_Q}\cdot d\mathbf{\dot{Q}} +\mathbf{P_A}\cdot d\mathbf{%
\dot{A}},  \label{thets}
\end{eqnarray}
where the difference $\vartheta_{2}-\vartheta_{1}$ is the exact one-form $%
d\left( \mathbf{P_Q\cdot\dot{Q}+P_A\cdot\dot{A}}\right) $ on $TT^{\ast}AQ$.

Let us note that, with the triplet (\ref{TAcc}), we have achieved to recast
the second order Lagrangian formalism as a Lagrangian submanifold given by
\begin{equation}
\left( T\pi_{ AQ}\right) ^{\ast}dL_{2}=\vartheta_{2},  \label{2ndELLag}
\end{equation}
where $T\pi_{ AQ}$ is the tangent mapping and the superscript denotes the
pull-back operation. In the local chart, this Lagrangian submanifold can be
written as
\begin{equation*}
\frac{\partial L_2}{\partial \mathbf{Q}}=\mathbf{\dot{P}_Q}, \qquad \frac{%
\partial L_2}{\partial \mathbf{A}}=\mathbf{\dot{P}_A}, \qquad \frac{\partial
L_2}{\partial \mathbf{\dot{Q}}}=\mathbf{P_Q}, \qquad \frac{\partial L_2}{%
\partial \mathbf{\dot{A}}}=\mathbf{P_A}.
\end{equation*}
Note that, by combining the first and third, combining the second and the
fourth, we arrive at the Euler-Lagrange equations (\ref{ELtwo}) which covers
the second order Euler-Lagrange equations under the assumption that (\ref%
{chi}) is satisfied and the matrix $[\partial^{2}F/\partial\mathbf{\mathbf{A}%
}\partial\mathbf{\mathbf{\dot{Q}}]}$ is non degenerate.

Recalling the potential function, we define the energy function as follows
\begin{equation}  \label{HMF}
E^{L\rightarrow H}\left( \mathbf{Q,A,P_Q,P_A,\dot{Q},\dot{A}}\right) =
\mathbf{P_Q\cdot\dot{Q}+P_A\cdot\dot{A}}-L_2\left(\mathbf{Q,A,\dot{Q},\dot{A}%
}\right) ,
\end{equation}
satisfies the requirements, given in Eq.(\ref{MorseReq}), of being a Morse
family on the Pontryagin bundle $TAQ\times T^{\ast}AQ$ over the cotangent
bundle $T^{\ast}AQ$. Hence, $E^{L\rightarrow H}$ generates a Lagrangian
submanifold $\mathcal{S}_{T^{\ast}T^{\ast}\mathcal{Q}}$ of $%
T^{\ast}T^{\ast}Q $ as defined in Eq.(\ref{LagSub}). In coordinates
\begin{equation*}
\left( \mathbf{Q,A,P_Q,P_A,\Pi_Q,\Pi_A,\Phi_Q,\Phi_A}\right )\in
T^{\ast}T^{\ast}AQ
\end{equation*}
the submanifold $\mathcal{S}_{T^{\ast }T^{\ast}Q}$ is given by
\begin{eqnarray}
\mathbf{\Pi_Q}&=&\frac{\partial E^{L\rightarrow H}}{\partial \mathbf{Q}}=-%
\frac{\partial L_2}{\partial \mathbf{Q}},\qquad \mathbf{\Pi_A}=\frac{%
\partial E^{L\rightarrow H}}{\partial \mathbf{A}}=-\frac{\partial L_2}{%
\partial \mathbf{A}} \\
\mathbf{\Phi_Q}&=&\frac{\partial E^{L\rightarrow H}}{\partial \mathbf{P_Q}}=
\mathbf{\dot{Q}}, \qquad \mathbf{\Phi_A}=\frac{\partial E^{L\rightarrow H}}{%
\partial \mathbf{P_A}}= \mathbf{\dot{A}} \\
\mathbf{0}&=&\frac{\partial E^{L\rightarrow H}}{\partial \mathbf{\dot{Q}}}=%
\mathbf{P_Q}-\frac{\partial L_2}{\partial \mathbf{\dot{Q}}}, \qquad \mathbf{0%
}=\frac{\partial E^{L\rightarrow H}}{\partial \mathbf{\dot{A}}}=\mathbf{P_A}-%
\frac{\partial L_2}{\partial \mathbf{\dot{A}}}  \label{Morsefam}
\end{eqnarray}
The inverse musical isomorphism $\Omega_{T^{\ast}Q}^{\sharp}$ maps this
Lagrangian submanifold $\mathcal{S}_{T^{\ast}T^{\ast}Q}$ to the one defined
in \ref{2ndELLag} generated by the Lagrangian $L_2$. The last two equations
define the associated canonical momenta
\begin{align}
\mathbf{P}_{Q} & =\frac{\partial L_{2}}{\partial\mathbf{\mathbf{\mathbf{%
\dot {Q}}}}}=\frac{\partial{L}}{\partial\mathbf{\mathbf{\mathbf{\dot{Q}}}}}+%
\frac{\partial F}{\partial\mathbf{Q}}+\frac{\partial^{2}F}{\partial \mathbf{%
\mathbf{\dot{Q}}\partial Q}}\cdot\mathbf{\mathbf{\dot{Q}}+}\frac{%
\partial^{2}F}{\partial\mathbf{\mathbf{\dot{Q}}\partial\mathbf{\dot{Q}}}}%
\cdot\mathbf{A+}\frac{\partial F}{\mathbf{\partial\mathbf{\dot{Q}}}\partial%
\mathbf{\mathbf{A}}}\cdot\mathbf{\dot{A}}  \label{P-1} \\
\mathbf{P}_{A} & =\frac{\partial L_{2}}{\partial\mathbf{\dot{A}}}=\frac{%
\partial F}{\partial\mathbf{A}}\left( \mathbf{Q},\mathbf{\dot{Q}},\mathbf{A}%
\right) .  \label{P-2}
\end{align}
After the substitution of (\ref{P-2}), the Hamiltonian Morse family (\ref%
{HMF}) on the Whitney product space $T^{\ast} AQ\times_{ AQ}T AQ $ over $%
T^{\ast} AQ$ becomes free of $\mathbf{\dot{A}}$ and turns out to be
\begin{equation*}
H\left( \mathbf{Q,A,\mathbf{\mathbf{\dot{Q},}}P}_{Q},\mathbf{P}_{A}\right) =%
\mathbf{P}_{Q}\cdot\mathbf{\dot{Q}}-L\left( \mathbf{Q,\mathbf{\mathbf{\dot {Q%
}},}A}\right) -\frac{\partial F}{\partial\mathbf{Q}}\cdot \mathbf{\mathbf{%
\dot{Q}}-}\frac{\partial F}{\partial\mathbf{\mathbf{\dot{Q}}}}\cdot\mathbf{A.%
}
\end{equation*}
This is an example of a Morse family reduction.

If we assume that the matrix $[\partial^{2}F/\partial\mathbf{A}\partial%
\mathbf{\mathbf{\dot{Q}}]}$ is non-degenerate, from the equation (\ref{P-2}%
), we can solve $\mathbf{\dot{Q}}$ in terms of the momenta that is
\begin{equation}
\mathbf{\dot{Q}}=\mathbf{Z}\left( \mathbf{Q},\mathbf{A,P}_{A}\right) .
\label{Z}
\end{equation}
This enables us to arrive the isomorphic copies of the functions on $T AQ$
depending only on the variables $\left( \mathbf{Q},\mathbf{\dot{Q}},\mathbf{%
\mathbf{A}}\right) $ on the space $T^{\ast } AQ$ by the direct substitution.
So that, futher reduction of the Morse family is possible given by
\begin{equation}
H\left( \mathbf{Q,A,P}_{Q},\mathbf{P}_{A}\right) =\mathbf{P}_{Q}\cdot\mathbf{%
Z}-L\left( \mathbf{Q,Z,A}\right) -\frac{\partial F}{\partial\mathbf{Q}}\cdot%
\mathbf{Z-}\frac{\partial F}{\partial\mathbf{Z}}\cdot\mathbf{A}  \label{canH}
\end{equation}
can be defined on the cotangent bundle $T^{\ast} AQ$. In this case, the
Hamiltonian function $H$ on $T^{\ast} AQ$ determines the Lagrangian
submanifold $\mathcal{S}_{T^{\ast }T^{\ast}Q}$ in (\ref{Morsefam}) by means
of the Hamilton's equations
\begin{equation}
-\left( \tau_{T^{\ast} AQ}\right) ^{\ast}dH=\vartheta_{1},  \label{2ndHam}
\end{equation}
where the Hamiltonian function $H$ is the one in (\ref{canH}) and $%
\vartheta_1$ is in Eq.(\ref{thets}). Locally, the Hamilton's equations are
\begin{equation}
\mathbf{\dot{Q}}=\frac{\partial H}{\partial\mathbf{P}_{Q}},\qquad \mathbf{%
\dot{A}}=\frac{\partial H}{\partial\mathbf{P}_{A}}, \qquad\mathbf{\dot {P}}%
_{Q}=-\frac{\partial H}{\partial\mathbf{Q}}, \qquad\mathbf{\dot{P}}_{A}=-%
\frac{\partial H}{\partial\mathbf{A}}.  \label{HamEqs}
\end{equation}
The first equations are the defining identity (\ref{Z}), the second is
identically satisfied after the substitutions of the transformations (\ref%
{P-1} and \ref{P-2}), whereas the third and fourth ones are the
Euler-Lagrange equations.

In order to generalize the present discussion to any even order Lagrangian
density, one simply needs to replace $AQ$ with the higher order bundle $A^kQ$%
. In this case, simple manipulations and presentation some indices will be
enough to achieve this.

\subsection{Schmidt-Legendre transformation for odd orders}

\label{tol}

For a third order Lagrangian theory ${L}\left( \mathbf{q;\mathbf{\dot{q};}%
\ddot{q};\dddot{q}}\right)$, to link the acceleration $\mathbf{A}$ with the
derivative of the velocity $\mathbf{\dot{Q}}$, we need to introduce a
trivial bundle $T AQ\times TM$ over the tangent bundle $T AQ$. Here, $M$ is
an $n-$dimensional manifold with local coordinates $(\mathbf{r})$. Using the
isomorphism (\ref{iso}), we define a Morse family
\begin{equation}
L_{3}\left( \mathbf{Q},\mathbf{\mathbf{A};\mathbf{\dot{Q}},\mathbf{\dot {A}%
,r,}\dot{r}}\right) ={L}\left( \mathbf{Q},\mathbf{\dot{Q}},\mathbf{\mathbf{A,%
\dot{A}}}\right) +\frac{\partial F}{\partial\mathbf{Q}}\cdot\mathbf{\mathbf{%
\dot{Q}}+}\frac{\partial F}{\partial\mathbf{\mathbf{\dot {Q}}}}\cdot\mathbf{%
\mathbf{A}}+\frac{\partial F}{\partial\mathbf{\mathbf{A}}}\cdot\mathbf{%
\mathbf{\dot{A}+}}\frac{\partial F}{\partial\mathbf{\mathbf{r}}}\cdot\mathbf{%
\dot{r}.}  \label{L3}
\end{equation}
on the total space $T AQ\times TM$. In this case, the auxiliary function $F$
depends on the base components of the manifold $M$ as well, that is $F=F(
\mathbf{Q},\mathbf{A},\mathbf{\dot{Q}},\mathbf{r})$ is defined on $%
AQ\times_Q TQ \times M$. over the base manifold $T AQ$ which determines a
Lagrangian submanifold of $T^{\ast }T AQ$. A simple calculations shows that,
the Morse family $L_{3}$ generates the 3rd order Euler-Lagrange equations
only if the single requirement
\begin{equation}
det[\partial^{2}F/\partial\mathbf{\mathbf{\mathbf{\dot{Q}}}\partial r}]\neq0
\label{cond2}
\end{equation}
is employed.

In the classical sense, canonical momenta for the cotangent bundle $T^{\ast
}( AQ\times M)$ are%
\begin{align}
\mathbf{P}_{Q} & =\frac{\partial L_{3}}{\partial\mathbf{\mathbf{\mathbf{%
\dot {Q}}}}}=\frac{\partial{L}}{\partial\mathbf{\mathbf{\mathbf{\dot{Q}}}}}+%
\frac{\partial F}{\partial\mathbf{Q}}+\frac{\partial^{2}F}{\partial \mathbf{%
\mathbf{\dot{Q}}\partial Q}}\cdot\mathbf{\mathbf{\dot{Q}}+}\frac{%
\partial^{2}F}{\partial\mathbf{\mathbf{\dot{Q}}\partial\mathbf{\dot{Q}}}}%
\cdot\mathbf{A+}\frac{\partial F}{\mathbf{\partial\mathbf{\dot{Q}}}\partial%
\mathbf{\mathbf{A}}}\cdot\mathbf{\dot{A}+}\frac{\partial F}{\mathbf{\partial%
\mathbf{\dot{Q}}}\partial\mathbf{\mathbf{r}}}\cdot\mathbf{\dot{r},}
\label{mom3rd} \\
\mathbf{P}_{A} & =\frac{\partial L_{3}}{\partial\mathbf{\dot{A}}}=\frac{%
\partial{L}}{\partial\mathbf{\dot{A}}}+\frac{\partial F}{\partial \mathbf{A}}%
,  \notag \\
\mathbf{P}_{r} & =\frac{\partial L_{3}}{\partial\mathbf{\dot{r}}}=\frac{%
\partial F}{\partial\mathbf{r}},  \notag
\end{align}
where $\left( \mathbf{P}_{Q},\mathbf{P}_{A},\mathbf{P}_{r}\right) $ is the
conjugate fiber coordinates. If the non-degeneracy of $[\partial^{2}F/%
\partial\mathbf{\dot{Q}\partial r}]$ is assumed, it is possible to solve $%
\mathbf{\dot{Q}}$ in terms of $\left( \mathbf{Q},\mathbf{\mathbf{A,r,}P}%
_{r}\right) $. In this case, we write $\mathbf{\dot{Q}}=\mathbf{W}\left(
\mathbf{Q},\mathbf{\mathbf{A,r,}P}_{r}\right)$. Note that, the third order
Lagrangian density $L$ is no need to be non-degenerate in the sense of
Ostragradsky.

Now we assume particularly that the third order Lagrangian ${L}$ that we
started with depends only on $(\mathbf{q},\mathbf{\dot{q}},\mathbf{\ddot{q}}%
) $. That is, it actually is a second order Lagrangian. In this case, the
Lagrangian $L_{3}$ becomes
\begin{equation}
L_{3}\left( \mathbf{Q},\mathbf{\mathbf{A};\mathbf{\dot{Q}},\mathbf{\dot {A}%
,r,}\dot{r}}\right) ={L}\left( \mathbf{Q},\mathbf{\dot{Q}},\mathbf{\mathbf{A}%
}\right) +\frac{\partial F}{\partial\mathbf{Q}}\cdot\mathbf{\mathbf{\dot{Q}}+%
}\frac{\partial F}{\partial\mathbf{\mathbf{\dot {Q}}}}\cdot\mathbf{\mathbf{A}%
}+\frac{\partial F}{\partial\mathbf{\mathbf{r}}}\cdot\mathbf{\dot{r},}
\label{L-3}
\end{equation}
where we assume $F=F(\mathbf{Q,\mathbf{\dot{Q},r}})$. This time the
canonical momenta (\ref{mom3rd}) take the particular form
\begin{align}
\mathbf{P}_{Q} & =\frac{\partial L_{3}}{\partial\mathbf{\mathbf{\mathbf{%
\dot {Q}}}}}=\frac{\partial{L}}{\partial\mathbf{\mathbf{\mathbf{\dot{Q}}}}}+%
\frac{\partial F}{\partial\mathbf{Q}}+\frac{\partial^{2}F}{\partial \mathbf{%
\mathbf{\dot{Q}}\partial Q}}\cdot\mathbf{\mathbf{\dot{Q}}+}\frac{%
\partial^{2}F}{\partial\mathbf{\mathbf{\dot{Q}}\partial\mathbf{\dot{Q}}}}%
\cdot\mathbf{A+}\frac{\partial F}{\mathbf{\partial\mathbf{\dot{Q}}}\partial%
\mathbf{\mathbf{r}}}\cdot\mathbf{\dot{r}}  \label{canmom3} \\
\mathbf{P}_{A} & =\frac{\partial L_{3}}{\partial\mathbf{\dot{A}}}=\mathbf{0},
\\
\mathbf{P}_{r} & =\frac{\partial L_{3}}{\partial\mathbf{\dot{r}}}=\frac{%
\partial F}{\partial\mathbf{r}},
\end{align}
where $\mathbf{P}_{A}=\mathbf{0}$ is a set of primary constraints. The
non-degeneracy of $[\partial^{2}F/\partial\mathbf{\mathbf{\mathbf{\dot{Q}}}%
\partial r}]$ enough to write $\mathbf{\mathbf{\mathbf{\dot{Q}}}}$ and $%
\mathbf{\dot{r}}$ in terms of $\left( \mathbf{Q},\mathbf{\mathbf{A,r,\mathbf{%
P}_{Q},}P}_{r}\right) $. The total Hamiltonian is given by
\begin{equation}
H_{T}=\mathbf{P}_{Q}\cdot\mathbf{\mathbf{\mathbf{\dot{Q}-L-}}}\frac{\partial
F}{\partial\mathbf{Q}}\cdot\mathbf{\mathbf{\dot{Q}-\frac{\partial F}{\partial%
\mathbf{\mathbf{\dot{Q}}}}\cdot\mathbf{\mathbf{A}}+\lambda}}\cdot\mathbf{P}%
_{A},  \label{totHam}
\end{equation}
where $\mathbf{\mathbf{\lambda}}$ is the vector of Lagrange multipliers.
Geometrically, by employing the constraint $\Phi_{1}=\mathbf{P}_{A}=0$, the
total Hamiltonian can be reduced to $\left( T^{\ast}Q\times_{Q} AQ\right)
\times T^{\ast}M.$ But further reduction on the total space is needed for
the consistency of the primary constraint $\Phi_{1}$. To check the
consistency of $\Phi_{1}=\mathbf{P}_{A}$, we take the Poisson bracket of $%
\Phi_{1}$ and $H_{T}$ which gives that, the time derivative
\begin{equation*}
\Phi_{2}=\dot{\Phi}_{1}=\left\{ \mathbf{P}_{A},H_{T}\right\} =\frac {\partial%
{L}}{\partial\mathbf{A}}+\frac{\partial{F}}{\partial \mathbf{\mathbf{\dot{Q}}%
}}\approx\mathbf{0}
\end{equation*}
must be weakly zero. To check the consistency of $\Phi_{2},$ we compute
\begin{equation*}
\dot{\Phi}_{2}=\left\{ \Phi_{2},H_{T}\right\}
\end{equation*}
which equals to
\begin{equation*}
\frac{\partial^{2}{L}}{\partial\mathbf{A}^{2}}\cdot\mathbf{\mathbf{\lambda+}}%
\frac{\partial^{2}{L}}{\partial\mathbf{Q}^{2}}\cdot\mathbf{\mathbf{\dot{Q}+}}%
\frac{\partial^{2}{L}}{\partial\mathbf{Q}\partial\mathbf{\mathbf{\dot{Q}}}}%
\cdot\mathbf{A+P}_{Q}-\frac{\partial{L}}{\partial\mathbf{\mathbf{\dot{Q}}}}-%
\frac{\partial{F}}{\partial\mathbf{\mathbf{Q}}}=0.
\end{equation*}
If the Lagrangian is assumed to be non-degenerate then this step determines
the Lagrange multipliers $\mathbf{\mathbf{\lambda}}$, and the constraint
algorithm is finished up. If the Lagrangian is degenerate further steps may
be needed to determine the Lagrange multipliers as well as to close up the
Poisson algebra. We refer \cite{AnGoMaMa10} for the present and further
discussions on this.

It is immediate to generalize this for the higher order theories.

\section{Symplectic relation between Schmidt and Ostragradsky}

\subsection{For even order formalisms}

For a second order Lagrangian density on $T^{2}Q$, the corresponding
Hamiltonian formulation obtained after performing Ostrogradsky Legendre
transformation is on the canonical symplectic space $T^{\ast}TQ$, whereas
the corresponding Hamiltonian formulation obtained after performing Schmidt
Legendre transformation is on $T^{\ast} AQ$. Here, $AQ$ is the acceleration
bundle presented in the first section. In this section, we establish a
purely symplectic transformation between $T^{\ast}TQ$ and $T^{\ast} AQ$.

The symplectic transformation (relation) between two spaces can be obtained
by following \cite{Be11,Tu76,Tu77,Tu80}. To this end, we first recall the
following coordinates $\left( \mathbf{q}_{1}\mathbf{,q}_{2}\mathbf{;\pi}^{1},%
\mathbf{\pi}^{2}\right) $ on $T^{\ast}TQ$ and $\left( \mathbf{Q},\mathbf{%
\mathbf{A};\mathbf{P}}_{Q}\mathbf{,\mathbf{P}}_{A}\right) $ on $T^{\ast} AQ$%
. The canonical Liouville one-forms are given by
\begin{equation*}
\vartheta_{T^{\ast}TQ}=\mathbf{\pi}^{1}\cdot d\mathbf{q}_{1}+\mathbf{\pi}%
^{2}\cdot d\mathbf{q}_{2}\text{ \ \ and \ \ }\vartheta_{T^{\ast} AQ}=\mathbf{%
\mathbf{P}}_{Q}\cdot d\mathbf{Q}+\mathbf{\mathbf{P}}_{A}\cdot d\mathbf{%
\mathbf{A}}\text{,}
\end{equation*}
whose exterior derivatives $d\vartheta_{T^{\ast}TQ}$ and $%
d\vartheta_{T^{\ast }TQ}$ are symplectic forms on $T^{\ast}TQ$ and $T^{\ast}
AQ$. Recall that the momenta in Eq.(\ref{P-1} and \ref{P-2}) defines the
following isomorphism

\begin{equation*}
T^{\ast} AQ\rightarrow T AQ:\left( \mathbf{Q,A};\mathbf{P}_{Q},\mathbf{P}%
_{A}\right) \rightarrow\left( \mathbf{Q,A};\mathbf{Z}\left( \mathbf{Q,A,P}%
_{A}\right) ,\mathbf{\dot{A}}\right) ,
\end{equation*}
where $\mathbf{\dot{A}}$ is a smooth function of $\left( \mathbf{Q},\mathbf{%
\mathbf{A};\mathbf{P}}_{Q}\mathbf{,\mathbf{P}}_{A}\right) $. Composing the
canonical map $\hat{\kappa}_{Q}$ in (\ref{kappahat}) map, and the
acceleration bundle projection $\mathfrak{t}_{TQ}$ presented in (\ref{accT}%
), we arrive a surjective projection of $T^{\ast} AQ$ on $TQ$ given by
\begin{equation}
T^{\ast} AQ\rightarrow TQ:\left( \mathbf{Q,A};\mathbf{P}_{Q},\mathbf{P}%
_{A}\right) \rightarrow\left( \mathbf{Q;\mathbf{Z}\left( \mathbf{Q,A,P}%
_{A}\right) }\right) .  \label{proj}
\end{equation}
This mapping enables us to define a Whitney product
\begin{equation}
W=T^{\ast} AQ\times_{TQ}T^{\ast}TQ  \label{W}
\end{equation}
over the tangent bundle $TQ$, where the projection $T^{\ast} AQ\rightarrow TQ
$ in the one in Eq.(\ref{proj}) and $T^{\ast}TQ\rightarrow TQ$ is the
cotangent bundle projection. Hence, we can take coordinates on the Whitney
product $W$ as $\left( \mathbf{Q,\pi}^{1},\mathbf{\pi}^{2};\mathbf{A};%
\mathbf{P}_{Q},\mathbf{P}_{A}\right) $ by identifying $\mathbf{q=Q}$ and $%
\mathbf{\dot{Q}=\mathbf{Z}\left( \mathbf{Q,A,P}_{A}\right) }$, so that $W$
is $6n$-dimensional. By pulling the Liouville forms $\vartheta_{T^{\ast}TQ}$
and $\vartheta_{T^{\ast} AQ}$ back to $W,$ we arrive the following one form%
\begin{equation*}
\vartheta_{W}=\vartheta_{T^{\ast} AQ}\ominus\vartheta_{T^{\ast}TQ}=\mathbf{P}%
_{Q}d\mathbf{Q}+\mathbf{P}_{A}d\mathbf{A}-\mathbf{\pi}^{1}d\mathbf{Q}-%
\mathbf{\pi}^{2}d\mathbf{\mathbf{Z}}\text{.}
\end{equation*}
The exterior derivative of $\omega_{W}=d\vartheta_{W}$ is a symplectic
manifold on $W$.

A symplectic transformation (or a relation) between $T^{\ast} AQ$ and $%
T^{\ast}TQ$ is a Lagrangian submanifolds of $W$. A Lagrangian submanifold $S$
of $W$ has dimension $3n$ and the restriction of the symplectic form to $S$
vanishes. The total product space $W$ in (\ref{W}) can be understood as the
cotangent bundle $T^{\ast}T^{2}Q$ of the second order tangent bundle $T^{2}Q$
identified with the Whitney product $TQ\times_{Q} AQ$. In this case, the
exterior derivative of a function $F=F\left( \mathbf{Q},\mathbf{Z},\mathbf{%
\mathbf{A}}\right) $ on $T^{2}Q$ defines a Lagrangian submanifold $S.$ To
arrive the Legendre transformation, we equate the one-form $\vartheta_{W}$
with the exterior derivative of $F$, and this locally looks like
\begin{equation*}
\mathbf{P}_{Q}d\mathbf{Q}+\mathbf{P}_{A}d\mathbf{A}-\mathbf{\pi}^{1}d\mathbf{%
Q}-\mathbf{\pi}^{2}d\mathbf{Z}=\frac{\partial F}{\partial \mathbf{Q}}d%
\mathbf{Q+}\frac{\partial F}{\partial\mathbf{Z}}d\mathbf{Z+}\frac{\partial F%
}{\partial\mathbf{A}}d\mathbf{A.}
\end{equation*}
For this we arrive the transformation by the following list of relations
\begin{equation}
\mathbf{P}_{Q}-\mathbf{\pi}^{1}-\frac{\partial F}{\partial\mathbf{Q}}=0\text{%
, \ \ }\mathbf{P}_{A}-\frac{\partial F}{\partial\mathbf{A}}=0\text{, \ \ }%
\mathbf{\pi}^{2}-\frac{\partial F}{\partial\mathbf{Z}}=0  \label{Legtrf}
\end{equation}
where we choose\ $\mathbf{\dot{Q}=\mathbf{Z}\left( \mathbf{Q,A,P}_{A}\right)
}$. These equations are the Legendre transformation relating the Schmidt and
Ostrogradsky transformations \cite{AnGoMaMa10}. Explicitly, the symplectic
diffeomorphism is given by%
\begin{equation*}
T^{\ast} AQ\rightarrow T^{\ast}TQ:\left( \mathbf{Q},\mathbf{\mathbf{A};%
\mathbf{P}}_{Q}\mathbf{,\mathbf{P}}_{A}\right) \rightarrow\left( \mathbf{Q},%
\mathbf{Z}\left( \mathbf{Q},\mathbf{\mathbf{A}},\mathbf{\mathbf{P}}%
_{A}\right) ;\mathbf{\mathbf{P}}_{Q}-\frac{\partial F}{\partial\mathbf{Q}}%
\left( \mathbf{Q},\mathbf{Z}\left( \mathbf{Q},\mathbf{\mathbf{A}},\mathbf{%
\mathbf{P}}_{A}\right) ,\mathbf{\mathbf{A}}\right) ,-\frac{\partial F}{%
\partial\mathbf{Z}}\right) .
\end{equation*}

As a summary, we can say that the Gauge invariance of the Lagrangian is the
addition of a total derivative $dF/dt$ to the Lagrangian function. For
second order theories, $F$ is defined on $T^{2}Q$ identified with $TQ\times
_{Q} AQ$ so that its exterior derivative is a Lagrangian submanifold of $%
T^{\ast}T^{2}Q$ which enaples us to define the canonical diffeomorphism (\ref%
{Legtrf}) from $T^{\ast} AQ$ to $T^{\ast}TQ$.

\subsection{For the odd order formalisms}

Define two bundle structures by considering the base manifold as $AQ$,
namely the trivial bundle $AQ\times M$ and $T^2Q$. Here, the manifold $M$ is
the one presented in the subsection (\ref{tol}) whereas the bundle structure
of $T^2Q$ is the one described in (\ref{dia1}). Accordingly, we take the
Whitney product of these two bundle and arrive $T^2Q\times_{AQ}(AQ\times M)$
with local coordinates $(\mathbf{Q,A,\dot{Q},r})$.

The function $F=F(\mathbf{Q,A,\dot{Q},r})$ generating the gauge invariance
of the third order Lagrangian function is defined on $T^2Q\times_{AQ}(AQ%
\times M)$ hence the image space of its exterior derivative is a Lagrangian
submanifold of the cotangent bundle
\begin{equation}  \label{U}
U= T^*(T^2Q\times_{AQ}(AQ\times M))\simeq T^*T^2Q\times_{AQ}T^*(AQ\times M)).
\end{equation}
So that it generates a symplectic diffeomorphism between the cotangent
bundles $T^*T^2Q$ and $T^*(AQ\times M))$. Observe that, $T^*T^2Q$ is the
momentum phase space for the third order Lagrangian formalism when it is
transformed in the sense of Ostragradsky, and $T^*(AQ\times M))$ is the
momentum phase space for the third order Lagrangian formalism when it is
transformed in the sense of Schmidt.

Let us first introduce the coordinates $\left( \mathbf{q}_{1},\mathbf{q}_{2},%
\mathbf{q}_{3};\mathbf{\pi}^{1},\mathbf{\pi}^{2},\mathbf{\pi}^{3}\right) $
on $T^*T^2Q$ then the canonical one-forms given by
\begin{equation*}
\vartheta_{T^{\ast}T^2Q}=\mathbf{\pi}^{1}\cdot d\mathbf{q}_{1}+\mathbf{\pi}%
^{2}\cdot d\mathbf{q}_{2}+\mathbf{\pi}^{3}\cdot d\mathbf{q}_{3}, \qquad
\vartheta_{T^*(AQ\times M)}=\mathbf{\mathbf{P}}_{Q}\cdot d\mathbf{Q}+\mathbf{%
\mathbf{P}}_{A}\cdot d\mathbf{\mathbf{A}}+\mathbf{\mathbf{P}}_{r}\cdot d%
\mathbf{r}.
\end{equation*}
By pulling the one-forms $\vartheta_{T^{\ast}TQ}$ and $\vartheta_{T^{\ast}
AQ}$ back to $U$ we arrive at the following one form%
\begin{equation*}
\vartheta_{U}=\vartheta_{T^{\ast} AQ}\ominus\vartheta_{T^{\ast}TQ}=\mathbf{%
\mathbf{P}}_{Q}\cdot d\mathbf{Q}+\mathbf{\mathbf{P}}_{A}\cdot d\mathbf{%
\mathbf{A}}+\mathbf{\mathbf{P}}_{r}\cdot d\mathbf{\mathbf{r}}-\mathbf{\pi}%
^{1}d\mathbf{Q}-\mathbf{\pi}^{2}d\mathbf{\mathbf{W}}-\mathbf{\pi}^{3}d%
\mathbf{\mathbf{A}}
\end{equation*}
whose exterior derivative is a symplectic two-form on $U$. Here, we take $%
\mathbf{q_3=A}$ and $\mathbf{q_1=Q}$ to verify the definition of the Whitney
bundle $U$ in (\ref{U}). To arrive at the Legendre transformation, we equate
the one-form $\vartheta_{U}$ with the exterior derivative of $F$, and this
locally reads
\begin{equation}
\mathbf{P}_{Q}-\mathbf{\pi}^{1}-\frac{\partial F}{\partial\mathbf{Q}}%
=0,\qquad \mathbf{P}_{A}-\mathbf{\pi}^{3}-\frac{\partial F}{\partial\mathbf{A%
}}=0\text{, \ \ }\mathbf{\pi}^{2}-\frac{\partial F}{\partial\mathbf{W}}=0
\label{Legtrf}
\end{equation}
where we write $\mathbf{\dot{Q}=\mathbf{W}\left( \mathbf{Q,A,r,P}_{A}\right)
}$ by solving the third equation in (\ref{mom3rd}).

\section{Examples}

\subsection{Example 1}

Let $Q=%
\mathbb{R}
$ and consider the non-degenerate second order Lagrangian given by
\begin{equation}
L=\frac{1}{2}\ddot{q}^{2}+5q^{2}\dot{q}^{2}+q^{6}  \label{L1}
\end{equation}
on $T^{2}Q$ with coordinates $\left( q,\dot{q},\ddot{q}\right) $. The 2nd
order Euler-Lagrange equations (\ref{EL2}) yield the following fourth-order
equation
\begin{equation}
q^{(iv)}-10q^{2}\ddot{q}-10q\dot{q}^{2}+6q^{5}=0.
\end{equation}
On the coordinates $\left( q,a,\dot{q},\dot{a}\right) $ on $T AQ$,
integrating the condition (\ref{chi}), we arrive the auxiliary function $%
F(q,a,\dot{q},\dot{a})=-\dot{q}a+W(q,a),$ for $W$ being an arbitrary
function. The Lagrangian $L_{2}$ in (\ref{L2}) takes the particular form%
\begin{equation}
L_{2}(q,a,\dot{q},\dot{a})=-\frac{1}{2}a^{2}+5q^{2}\dot{q}^{2}+q^{6}-\dot {q}%
\dot{a}+\frac{d}{dt}W(q,a)
\end{equation}
It is obvious that the total derivative term may be neglected. Following (%
\ref{P-1}-\ref{P-2}), we define the canonical momenta as%
\begin{equation*}
p_{q}=10q^{2}\dot{q}-\dot{a}\text{, \ \ }p_{a}=-\dot{q},
\end{equation*}
then the canonical Hamiltonian function (\ref{canH}) on $T^{\ast} AQ $ with
coordinates given by
\begin{equation*}
H=p_{q}p_{a}+\frac{1}{2}a^{2}-5q^{2}p_{a}^{2}-q^{6}.
\end{equation*}
The Hamilton's equations (\ref{HamEqs})\ are%
\begin{equation*}
\dot{q}=p_{a},\text{ \ \ }\dot{a}=p_{q}-10q^{2}p_{a}\text{, \ \ }\dot{p}%
_{q}=-10qp_{a}^{2}-6q^{5},\text{ \ \ }\dot{p}_{a}=a.
\end{equation*}

The Ostrogradsky Legendre transformation of the second order Lagrangian (\ref%
{L1}) can be achieved by defining the momenta (\ref{OsMo}) as%
\begin{equation*}
\pi ^{1}=10q_{1}^{2}q_{2}-\dot{q}_{2},\text{ \ \ }\pi ^{2}=\dot{q}_{1},
\end{equation*}%
where we choose $q_{1}=q$ and $q_{2}=\dot{q}$. The canonical Hamiltonian (%
\ref{canH2}) turns out to be
\begin{equation*}
H=q_{2}\pi ^{1}-\frac{1}{2}\left( \pi ^{2}\right)
^{2}-5q_{1}^{2}q_{2}^{2}-q^{6}.
\end{equation*}%
The transformation between these two system is given by
\begin{equation}
q=q_{1},\;\;p_{q}=\pi ^{1},\;\;\ P_{1}=\pi ^{2},\;\;P_{2}=-q_{2},
\end{equation}%
whereas the generating function $F=-q_{2}a.$

\subsection{Example 2}

The Pais-Uhlenbeck fourth-order oscillator equation of motion, \textit{viz},
\begin{equation}
q^{(iv)}+(\omega_{1}^{2}+\omega_{2}^{2})\ddot{q}+\omega_{1}^{2}%
\omega_{2}^{2}q=0,  \label{PUeqn}
\end{equation}
which is a quantum mechanical prototype of a field theory containing both
second and fourth order derivative terms \cite{Ma07,Mo10,pu1950}.
Mechanically the Pais-Uhlenbeck oscillator can be realized as a two-degrees
of freedom spring-mass system
\begin{equation}
m\ddot{x_{1}}+2kx_{1}-kx_{2}=0,\qquad m\ddot{x_{2}}-kx_{1}+2kx_{2}=0,
\end{equation}
where $k$ is the stiffness constant. This can be expressed by the
Pais-Uhlenbeck fourth-order oscillator equation where $\omega_{1}=\sqrt {%
\frac{3k}{m}}$ and $\omega_{2}=\sqrt{\frac{k}{m}}$. The Lagrangian of the
coupled system is given by
\begin{equation}
L_{coupled}=\frac{m}{2}(\dot{x_{1}}^{2}+\dot{x_{2}}^{2})-\frac{k}{2}\big(%
x_{1}^{2}+x_{2}^{2}+(x_{2}-x_{1})^{2}\big)\equiv T-U.
\end{equation}
It is easy to show that the corresponding Hamiltonian $H=T+U\equiv E_{m.e.}$%
( mechanical energy) and satisfies $\frac{dH}{dt}=0$. The Pais-Uhlenbeck
equation (\ref{PUeqn}) is generated by a degenerate fourth order Lagrangian
\begin{equation}
L_{PU-alt}=-\big(q\ddddot{q}+(\omega_{1}^{2}+\omega_{2}^{2})\ddot{q}%
+\omega_{1}^{2}\omega_{2}^{2}\,q\big).
\end{equation}
In this paper, we consider the non-degenerate second order Lagrangian of the
Pais-Uhlenbeck oscillator given by
\begin{equation}
L_{PU}=\frac{1}{2}\ddot{q}^{2}-\frac{1}{2}(\omega_{1}^{2}+\omega_{2}^{2})%
\dot{q}^{2}+\frac{1}{2}\omega_{1}^{2}\omega_{2}^{2}q^{2}.  \label{LagPU}
\end{equation}

We consider $Q=%
\mathbb{R}
$ with coordinate $q,$ $T^{2}Q$ with coordinates $\left( q,\dot{q},\ddot {q}%
\right) $, $T AQ$ with coordinates $\left( q,a,\dot{q},\dot {a}\right) $,
and write the Lagrangian $L_{2}$ in (\ref{L2}) for the case of
Pais-Uhlenbeck Lagrangian $L_{PU}$ in (\ref{LagPU}) as follows
\begin{equation}
L_{2}=-\frac{1}{2}a^{2}-\frac{1}{2}(w_{1}^{2}+w_{2}^{2})\dot{q}^{2}-\dot {q}%
\dot{a}+\frac{1}{2}w_{1}^{2}w_{2}^{2}q^{2},
\end{equation}
where we choose $F\left( \dot{q},\dot{a}\right) =-\dot{q}a$ by solving the
defining equation (\ref{chi}). the canonical momenta (\ref{P-1}-\ref{P-2})
take the particular form
\begin{equation*}
p_{q}=-(w_{1}^{2}+w_{2}^{2})\dot{q}-\dot{a}\text{, \ \ }p_{a}=-\dot{q},
\end{equation*}
whereas the canonical Hamiltonian function (\ref{canH}) turns out to be
\begin{equation*}
H=-p_{q}p_{a}+\frac{1}{2}a^{2}+\frac{1}{2}(\omega_{1}^{2}+%
\omega_{2}^{2})p_{a}^{2}-\frac{1}{2}\omega_{1}^{2}\omega_{2}^{2}q^{2}.
\end{equation*}
In this case the motion is generated by
\begin{equation}
\dot{q}=-p_{a},\text{ \ \ }\dot{a}=-p_{q}+(w_{1}^{2}+w_{2}^{2})p_{a},\text{
\ \ }\dot{p}_{q}=w_{1}^{2}w_{2}^{2}q,\text{ \ \ }\dot{p}_{a}=-a.
\label{HamEqPU}
\end{equation}

It is interesting at this juncture to compare these equations with those
following from Ostrogradsky's formulation. For the latter the new
coordinates are chosen to be $q_{1}=q$ and $q_{2}=\dot{q}$ while the
corresponding momenta (\ref{OsMo}) are
\begin{equation*}
\pi^{2}=\dot{q}_{2}\text{, \ \ }\pi^{1}=-(w_{1}^{2}+w_{2}^{2})q_{2}-\ddot {q}%
_{2}.
\end{equation*}
The Hamiltonian in (\ref{canH2}) has the explicit form
\begin{equation*}
H=\frac{1}{2}\left( \pi^{2}\right) ^{2}+\pi^{1}q_{2}+\frac{1}{2}%
(w_{1}^{2}+w_{2}^{2})q_{2}^{2}-\frac{1}{2}w_{1}^{2}w_{2}^{2}q_{1}^{2}
\end{equation*}
and leads to the following equations:
\begin{equation}
\dot{q}_{1}=q_{2},\text{ \ \ }\dot{q}_{2}=\pi^{2},\text{ \ \ }\dot{\pi}%
^{1}=w_{1}^{2}w_{2}^{2}q_{1},\text{ \ \ }\dot{\pi}^{2}=-%
\pi^{1}-(w_{1}^{2}w_{2}^{2})q_{2}.
\end{equation}
We refer \cite{ChFaLiTo13}. Comparing these first-order equations with those
obtained earlier (\ref{HamEqPU}), we observe the following relations
\begin{equation}
Q_{1}=q_{1},\;\;P_{1}=\pi^{1},\;\;Q_{2}=\pi^{2},\;\;P_{2}=-q_{2},
\label{N25}
\end{equation}
which clearly defines a canonical transformation between the phase space
variables as a particular case of (\ref{Legtrf}). Here, the generation
function is $F=-q_{2}a$.

\subsection{Example 3}

Although it seems that the condition (\ref{chi}) is not so strong, it is
possible to find some counterexamples. Let $Q=%
\mathbb{R}
^{2}$ with coordinates $\left( x,y\right) $ and consider the non-degenerate
Lagrangian
\begin{equation}
L=\frac{1}{2}(\dot{x}\ddot{y}^{2}+\dot{y}\ddot{x}^{2}),  \label{CountLag}
\end{equation}
then the integrability condition (\ref{chi}) becomes
\begin{equation}
\frac{\partial F}{\partial\dot{x}}=\ddot{x}\dot{y},\text{ \ \ }\frac{%
\partial F}{\partial\dot{y}}=\dot{x}\ddot{y}.  \label{Intcod}
\end{equation}
Note that the partial derivatives do not commute, that is $\partial
^{2}F/\partial\dot{x}\partial\dot{y}$ is not equal to $\partial^{2}F/\partial%
\dot{y}\partial\dot{x}$. This shows that, there is no such auxilary function
which makes the transformation possible in terms of Schmidt. We denote $%
\mathbf{q}_{\left( 1\right) }=\left( x,y\right) ,$ $\mathbf{q}_{\left(
2\right) }=\left( \dot{x},\dot{y}\right) ,$ $\mathbf{\pi}^{1}=\left(
\pi_{x}^{1},\pi_{y}^{1}\right) $, and $\mathbf{\pi}^{2}=\left(
\pi_{x}^{2},\pi_{y}^{2}\right) $. In this notation, we have that
\begin{equation*}
\mathbf{\dot{q}}_{\left( 2\right) }=\left( \ddot{x},\ddot{y}\right) =\left(
\frac{\pi_{x}^{2}}{\dot{y}},\frac{\pi_{y}^{2}}{\dot{x}}\right) ,
\end{equation*}
then the canonical Hamiltonian (\ref{canH2}) becomes
\begin{equation}
H=\mathbf{\dot{q}}_{\left( 1\right) }\cdot\mathbf{\pi}^{1}+\mathbf{\dot{q}}%
_{\left( 2\right) }\cdot\mathbf{\pi}^{2}-L\left( \mathbf{q}_{\left( 1\right)
},\mathbf{q}_{\left( 2\right) }\mathbf{,\dot{q}}_{\left( 2\right) }\right) .
\end{equation}

Using the method for degenerate cases, the Legendre transformation of (\ref%
{CountLag}) is possible because the integrability condition (\ref{chi}) is
not forced. In order to achieve this goal, we take $Q=%
\mathbb{R}
^{2}$ with coordinates $\left( x,y\right) $, and $AQ$ with induced
coordinates $\left( x,y;a,b\right) $. The augmentation of $AQ$ is given by
the manifold $M=%
\mathbb{R}
^{2}$ with auxiliary variables $\left( r,s\right) $. So that, the tangent
bundle of the product manifold $T\left( AQ\times M\right) $ has the
coordinates
\begin{equation*}
\left( x,y,a,b,r,s;\dot{x},\dot{y},\dot{a},\dot{b},\dot{r},\dot{s}\right) .
\end{equation*}
Obeying the condition (\ref{cond2}), we introduce the auxiliary function
\begin{equation*}
F=\dot{x}r+bs,
\end{equation*}
then the canonical momenta, denoted by $\left(
p_{x},p_{y},p_{a},p_{b},p_{r},p_{s}\right) $ and defined by (\ref{canmom3}),
have the following expressions,%
\begin{equation*}
p_{x}=b^{2}/2+\dot{r}\text{, \ \ }p_{y}=a^{2}/2+\dot{s},\text{ \ \ }%
p_{a}=p_{b}=0,\text{ \ \ }p_{r}=\dot{x},\text{ \ \ }p_{s}=\dot{y}.
\end{equation*}
Accordingly, the total Hamiltonian $H_{T}$ in (\ref{totHam}) takes the
particular form
\begin{equation*}
H_{T}=p_{x}p_{r}+p_{y}p_{s}-\left( a^{2}/2\right) p_{s}-\left(
b^{2}/2\right) p_{r}-ra-sb+\lambda^{1}p_{a}+\lambda^{2}p_{b}
\end{equation*}
where the Lagrange multipliers are $\left( \lambda^{1},\lambda^{2}\right) $.
The consistency check for the primary constraints $p_{a}\approx0$ and $%
p_{b}\approx0$ result with the secondary constraints
\begin{equation*}
\left\{ p_{a},H_{T}\right\} =-ap_{s}-r\approx0,\text{ \ \ }\left\{
p_{b},H_{T}\right\} =-bp_{r}-s\approx0\text{.}
\end{equation*}
The consistency checks of the secondary constraints read the tertiary
constraints
\begin{equation*}
\left\{ ap_{s}+r,H_{T}\right\} \approx0,\text{ \ \ }\left\{
bp_{r}+s,H_{T}\right\} \approx0,
\end{equation*}
which defines the Lagrangian multipliers
\begin{equation*}
\lambda^{1}=\frac{1}{p_{s}}\left( \frac{b^{2}}{2}-p_{x}-ab\right) ,\text{ \
\ }\lambda^{2}=\frac{1}{p_{r}}\left( \frac{a^{2}}{2}-p_{y}-ab\right) .
\end{equation*}

\subsection{Example 4}

Take $Q=%
\mathbb{R}
^{6}$ with coordinates $\left( \mathbf{X},\mathbf{Y}\right) $ both of which
are three dimensional vectors. We consider a degenerate second order
Lagrangian density
\begin{equation}
L=\frac{1}{2}\left( \mathbf{\dot{X}}^{2}+\mathbf{\dot{Y}}^{2}\right) +%
\mathbf{\dot{Y}\cdot \ddot{X}}-\frac{1}{2}\left( \mathbf{X}^{2}+\mathbf{Y}%
^{2}\right)   \label{stlag}
\end{equation}%
introduced in \cite{st06} where an investigation on topologically massive
gravity was performed. The acceleration bundle $AQ=%
\mathbb{R}
^{12}$ is equipped with the induced coordinates $\left( \mathbf{X},\mathbf{%
Y;A,B}\right) $ and six number of auxilary variables $\left( \mathbf{r},%
\mathbf{s}\right) $ as the coordinates of the manifold $M$ are introduced.
We take the auxiliary function
\begin{equation*}
F=\mathbf{\dot{X}\cdot r+\dot{Y}\cdot s}
\end{equation*}%
obeying the condition (\ref{cond2}). So that, we are ready to define the
Lagrangian density (\ref{L-3}) in the present case as follows%
\begin{equation*}
L=\frac{1}{2}\left( \mathbf{\dot{X}}^{2}+\mathbf{\dot{Y}}^{2}\right) +%
\mathbf{\dot{Y}\cdot A}-\frac{1}{2}\left( \mathbf{X}^{2}+\mathbf{Y}%
^{2}\right) +\mathbf{\dot{X}\cdot \dot{r}+\dot{Y}\cdot \dot{s}+A\cdot
r+B\cdot s}
\end{equation*}%
on the tangent bundle $T(AQ\times M)$ with the induced fiber coordinates $%
\left( \mathbf{\dot{X}},\mathbf{\dot{Y},\dot{A},\dot{B},\dot{r}},\mathbf{%
\dot{s}}\right) $. The canonical momenta (\ref{canmom3}) are computed to be
\begin{equation*}
\mathbf{P}_{X}=\mathbf{\dot{X}+\dot{r}},\text{ \ \ }\mathbf{P}_{Y}=\mathbf{%
\dot{Y}+\dot{s}+A},\text{ \ \ }\mathbf{P}_{A}=\mathbf{P}_{B}=0,\text{ \ \ }%
\mathbf{P}_{r}=\mathbf{\dot{X}},\text{ \ \ }\mathbf{P}_{s}=\mathbf{\dot{Y}}%
\text{,}
\end{equation*}%
where $\left( \mathbf{P}_{X},\mathbf{P}_{Y},\mathbf{P}_{A},\mathbf{P}_{B};%
\mathbf{P}_{r},\mathbf{P}_{s}\right) $ are the fiber coordinates of the
product cotangent bundle $T^{\ast }AQ\times T^{\ast }M$. Accordingly, the
primary constraints are $\mathbf{P}_{A}\approx 0$ and $\mathbf{P}_{B}\approx
0$. In this case, the total Hamiltonian (\ref{totHam}) turns out to be
\begin{equation*}
H_{T}=\mathbf{P}_{X}\cdot \mathbf{P}_{r}+\mathbf{P}_{s}\cdot (\mathbf{P}_{Y}-%
\mathbf{A})-\frac{1}{2}\left( \mathbf{P}_{r}^{2}+\mathbf{P}_{s}^{2}\right) +%
\frac{1}{2}\left( \mathbf{X}^{2}+\mathbf{Y}^{2}\right) -\mathbf{r\cdot
A-s\cdot B+\lambda \cdot P}_{A}+\mathbf{\beta }\cdot \mathbf{P}_{B}
\end{equation*}%
where $\mathbf{\lambda }$ and $\beta $ are Lagrange multipliers. The
consistency of the primary constraints leads to the secondary constraints%
\begin{equation}
\left\{ \mathbf{P}_{A},H_{T}\right\} =\mathbf{P}_{s}+\mathbf{r}\approx
\mathbf{0},\text{ \ \ }\left\{ \mathbf{P}_{B},H_{T}\right\} =\mathbf{s}%
\approx \mathbf{0.}  \label{st-sec}
\end{equation}%
We check the compatibility of these secondary constraints as well. These
give the tertiary constraints
\begin{equation}
\left\{ \mathbf{P}_{s}+\mathbf{r},H_{T}\right\} =\mathbf{B+P}_{X}-\mathbf{P}%
_{r}\approx \mathbf{0},\text{ \ \ }\left\{ \mathbf{s},H_{T}\right\} =\mathbf{%
P}_{Y}-\mathbf{A-P}_{s}\approx \mathbf{0.}  \label{st-tri}
\end{equation}%
Note that, the secondary and tertiary constraints in (\ref{st-sec}) and (\ref%
{st-tri}) define the auxiliary variables $\left( \mathbf{r,s}\right) $\ and
their conjugate momenta $\left( \mathbf{P}_{r},\mathbf{P}_{s}\right) $ in
terms of the coordinates $T^{\ast }AQ$. Continuing in this way, we compute
the fourtiery constraints
\begin{equation*}
\left\{ \mathbf{B+P}_{X}-\mathbf{P}_{r},H_{T}\right\} =\mathbf{\beta }-%
\mathbf{X-A}\approx \mathbf{0},\text{ \ \ }\left\{ \mathbf{P}_{Y}-\mathbf{A-P%
}_{s},H_{T}\right\} =-\mathbf{Y-\lambda -B}\approx \mathbf{0}
\end{equation*}%
which determine the Lagrange multipliers
\begin{equation*}
\mathbf{\lambda =}-\mathbf{Y-B},\text{ \ \ }\mathbf{\beta }=\mathbf{X+A.}
\end{equation*}%

So we arrive the Hamiltonian formulation of the dynamics generated by the
Lagrangian density (\ref{stlag}). The reduced Hamiltonian is given by
\begin{equation*}
H_{T}=\mathbf{P}_{X}\cdot \mathbf{B+P}_{X}^{2}+\left( \mathbf{P}_{Y}-\mathbf{%
A}\right) ^{2}-\frac{1}{2}\left( \left( \mathbf{B+P}_{X}\right) ^{2}+\left(
\mathbf{P}_{Y}-\mathbf{A}\right) ^{2}\right) +\frac{1}{2}\left( \mathbf{X}%
^{2}+\mathbf{Y}^{2}\right) +\left( \mathbf{P}_{Y}-\mathbf{A}\right) \cdot
\mathbf{A}
\end{equation*}%
on the manifold $T^{\ast }AQ$ with the constraints $\mathbf{P}_{A}\approx 0$
and $\mathbf{P}_{B}\approx 0$ are employed.

\subsection{Example 5}

We now consider the degenerate second order Lagrangian density
\begin{equation}
L=-\frac{1}{2}\dot{\mathbf{X}}^{2}+\mathbf{X}\cdot \left( \dot{\mathbf{X}}%
\times \ddot{\mathbf{X}}\right)   \label{Clelag}
\end{equation}%
which is similar to the one defined in \cite{cle94b}. Here, the inner
product $X^{2}=T^{2}-X^{2}-Y^{2}$ is defined by the Lorentzian metric and
the triple product is $\mathbf{X}\cdot (\mathbf{\dot{X}}\times \mathbf{\ddot{%
X}})=\epsilon _{ijk}X^{i}\dot{X}^{j}\ddot{X}^{k}$ where $\epsilon _{ijk}$ is
the completely antisymmetric tensor of rank three. We consider the base
manifold $Q=%
\mathbb{R}
^{3}$ with coordinates $\left( \mathbf{X}\right) $, the acceleration bundle $%
AQ=%
\mathbb{R}
^{6}$ with coordinates $\left( \mathbf{X,A}\right) $, and we consider the
auxiliary variables $\left( \mathbf{r}\right) $ as a coordinate frame of a
manifold $M=%
\mathbb{R}
^{3}$. We introduce the auxiliary function $F=\mathbf{\dot{X}\cdot r}$ which
enables us to write the Lagrangian density (\ref{L-3}) as%
\begin{equation*}
L=-\frac{1}{2}\dot{\mathbf{X}}^{2}+\mathbf{X}\cdot \left( \dot{\mathbf{X}}%
\times \mathbf{A}\right) +\mathbf{\dot{X}\cdot \dot{r}+r\cdot A}
\end{equation*}%
on $T(AQ\times M)$ with coordinates $\left( \mathbf{X,A,r;\dot{X},\dot{A},%
\dot{r}}\right) $. The conjugate coordinates on the cotangent bundle $%
T^{\ast }(AQ\times M)$ are $\left( \mathbf{X,A,r;P}_{X},\mathbf{P}_{A},%
\mathbf{P}_{r}\right) $ and following (\ref{canmom3}) they are defined by
\begin{equation*}
\mathbf{P}_{X}=-\dot{\mathbf{X}}+\mathbf{A}\times \mathbf{X+\dot{r},}\text{
\ \ }\mathbf{P}_{A}=\mathbf{0},\text{ \ \ }\mathbf{P}_{r}=\mathbf{\dot{X}}%
\text{. }
\end{equation*}%
This gives that $\mathbf{P}_{A}\approx \mathbf{0}$ are the primary
constraints. The total Hamiltonian (\ref{totHam}) becomes
\begin{equation}
H_{T}=\mathbf{P}_{X}\cdot \mathbf{P}_{r}+\frac{1}{2}\mathbf{P}_{r}^{2}-%
\mathbf{X}\cdot \left( \mathbf{P}_{r}\times \mathbf{A}\right) -\mathbf{r}%
\cdot \mathbf{\mathbf{\mathbf{\mathbf{A}}}}+\mathbf{\mathbf{\lambda }}\cdot
\mathbf{P}_{A},  \label{HT-C}
\end{equation}%
where $\mathbf{\mathbf{\lambda }}$ are the Lagrange multipliers.

Let us now check the consistency of the primary constraints
\begin{equation}
\left\{ \mathbf{P}_{A},H_{T}\right\} =\mathbf{P}_{r}\times \mathbf{X}-%
\mathbf{r}\approx \mathbf{0}  \label{Sc-C}
\end{equation}%
which determine a set of secondary constriants. The consistency check of
these secondary constraints (\ref{Sc-C}) lead to the tertiary constraints
\begin{equation*}
\left\{ \mathbf{P}_{r}\times \mathbf{X}-\mathbf{r},H_{T}\right\} =2\mathbf{%
\mathbf{\mathbf{\mathbf{A}}}}\times \mathbf{X}-\mathbf{P}_{r}-\mathbf{P}%
_{X}\approx \mathbf{0.}
\end{equation*}%
We check the compatibility of these constraints as well. These give the
fourtiery constriants
\begin{equation}
\left\{ 2\mathbf{\mathbf{\mathbf{\mathbf{A}}}}\times \mathbf{X}-\mathbf{P}%
_{r}-\mathbf{P}_{X},H_{T}\right\} =2\mathbf{\mathbf{\lambda }}\times \mathbf{%
X}+3\mathbf{\mathbf{\mathbf{\mathbf{A}}}}\times \mathbf{P}_{r}-\mathbf{%
\mathbf{\mathbf{\mathbf{A}}}}\approx \mathbf{0.}  \label{Cle-tri}
\end{equation}%
From these last equations, while solving $\mathbf{\mathbf{\lambda }}$, only
two of these three equations are independent, so that two components of the
Lagrange multiplier $\mathbf{\mathbf{\lambda }}$ can be determined depending
on the third one. By taking the dot product of (\ref{Cle-tri}) with $\mathbf{%
X}$ we arrive at a new constraint%
\begin{equation*}
3\mathbf{\mathbf{\mathbf{\mathbf{A}}}}\times \mathbf{P}_{r}\cdot \mathbf{X-%
\mathbf{\mathbf{\mathbf{A}}}}\cdot \mathbf{X}\approx \mathbf{0}.
\end{equation*}%
The consistency condition for this scalar constraint
\begin{equation*}
\left\{ 3\left( \mathbf{\mathbf{\mathbf{\mathbf{A}}}}\times \mathbf{P}%
_{r}\right) \cdot \mathbf{X-\mathbf{\mathbf{\mathbf{A}}}}\cdot \mathbf{X}%
,H_{T}\right\} =3\left( \mathbf{\mathbf{\lambda }}\times \mathbf{P}%
_{r}\right) \cdot \mathbf{X-\mathbf{\lambda }}\cdot \mathbf{X}-\mathbf{A}%
\cdot \mathbf{P}_{r}\approx \mathbf{0}
\end{equation*}%
determines the third and last component of the Lagrange multiplier $\mathbf{%
\mathbf{\lambda }}$. Substitution of the Lagrange multiplier result with the
explicit determination of the total Hamiltonian (\ref{HT-C}). We refer \cite%
{EsGuUc16} for the application of the Ostrogradsky-Legendre transformation
to the Sar\i oğlu-Tekin Lagrangian (\ref{stlag}) and Clemént Lagrangian (\ref%
{Clelag}).

\section{Conclusions \& Future work}

In this paper, we have constructed Tulczyjew triplet for the case of
acceleration bundle. This allowed us to study the higher order Lagrangian
formalism in terms of Lagrangian submanifolds and symplectic
diffeomorphisms. We have presented the symplectic relation between
Ostrogradsky and Schmidt methods of Legendre transformation. Several
examples both for the degenerate Lagrangians and the non degenerate
Lagrangians have been presented.

Some possible future works:

\begin{itemize}
\item To construct a unified formalism for the Schmidt-Legendre
transformation in terms Skinner-Rusk \cite{Sk83,SkRu83,SkRu83b}. This was
done for the case of Ostrogradsky-Legendre transformation in \cite{PrRo11},
and for the Hamilton-Jacobi theory \cite{CodeLePrRo}.

\item To study the Schmidt-Legendre transformation when the configuration
space is a Lie group and under the existence of some symmetries. We cite
\cite{CoDe11,GaHoMeRa12,GaHoRa11} for the Ostrogradsky-Legendre
transformation on Lie groups, and Ostrogradsky-Lie-Poisson reduction.
\end{itemize}

\section{Acknowledgement}

One of us (OE) is grateful to Hasan G{\"u}mral for pointing out the examples
of Sar\i oğlu-Tekin and Clem\'ent Lagrangians, and for many discussions on
this subject.

\bigskip

\bigskip

\bigskip

\bigskip

\end{document}